\documentclass{aa}  
\usepackage{xcolor}
\usepackage[normalem]{ulem}
\usepackage{graphicx}
\usepackage{placeins}
\usepackage{amsmath}
\usepackage{txfonts}
\usepackage{hyperref}
\usepackage{orcidlink}

\hypersetup{
colorlinks = true,
linkcolor = blue,
filecolor = blue,
citecolor = blue, 
urlcolor = blue}

\newcommand{\lfs}{\lambda_{\mathrm{fs}}}
\newcommand{\ev}{\mbox{eV}}
\newcommand{\kpch}{\mbox{$h^{-1}$ kpc}}
\newcommand{\mpch}
{\mbox{$h^{-1}$ Mpc}}

\newcommand{\gpch}{\mbox{$h^{-1}$ Gpc}}
\newcommand{\gpchc}{\mbox{$h^{-3}$ Gpc$^3$}}
\newcommand{\dif}{\mathrm{d}}
\newcommand{\leg}{\mathcal{L}}
\newcommand{\mnup}{\texttt{Mnu\_p}}
\newcommand{\mnupp}{\texttt{Mnu\_pp}}
\newcommand{\mnuppp}{\texttt{Mnu\_ppp}}
\newcommand{\sigmap}{\texttt{s8\_p}}
\newcommand{\sigmam}{\texttt{s8\_m}}
\newcommand{\fidza}{\texttt{fiducial\_ZA}}
\newcommand{\fid}{\texttt{fiducial}}
\newcommand{\chiq}{\chi^2}
\newcommand{\redchiq}{\tilde{\chi}^2}
\newcommand{\detec}{\mathrm{DET}}

\begin{document} 

   \title{The imprints of massive neutrinos on the three-point correlation function of large-scale structures}

   \author{Andrea Labate
          \inst{1,2}
          \fnmsep\thanks{\email{andrea.labate4@unibo.it}}
          \orcidlink{0009-0000-1002-9374}
          \and
          Massimo Guidi \inst{1,2}
          \orcidlink{0000-0001-9408-1101}
          \and
          Michele Moresco \inst{1,2}
          \orcidlink{0000-0002-7616-7136}
          \and 
          Alfonso Veropalumbo \inst{3,4,5}
          \orcidlink{0000-0003-2387-1194}
          }

   \institute{
   Dipartimento di Fisica e Astronomia ``Augusto Righi'', Universit\`a di Bologna, Via Piero Gobetti 93/2, I-40129 Bologna, Italy
   \and
    INAF–Osservatorio di Astrofisica e Scienza dello Spazio di Bologna, Via Piero Gobetti 93/3, I-40129 Bologna, Italy 
   \and 
    INAF-Osservatorio Astronomico di Brera, Via Brera 28, I-20122 Milano, Italy
   \and 
    INFN-Sezione di Genova, Via Dodecaneso 33, I-16146, Genova, Italy
   \and
   Dipartimento di Fisica, Universit\`a di Genova, Via Dodecaneso 33, I-16146, Genova, Italy
    }
    \date{}
 
  \abstract
   {}
   {Free-streaming of cosmic neutrinos affects the distribution and growth of cosmic structures on small scales. This enables the sum of neutrino masses $M_\nu$ to be constrained from clustering studies. 
    We investigate the possibility of disentangling massive neutrino cosmologies with the three-point correlation function (3PCF) for the first time. }
   {We measured the isotropic connected 3PCF $\zeta$ and the reduced 3PCF $Q$ of halo catalogs from the \textsc{Quijote} suite of $N$-body simulations, considering $M_\nu =0.0,  0.1, 0.2,$ and $0.4 \, \mathrm{eV}$ in different redshift bins.
    We developed a framework to quantify the detectability of massive neutrinos for different triangle configurations and shapes, and applied it to a case compatible with a stage-IV spectroscopic survey. We also compared our results with the analysis of simulations without neutrinos, but with different $\sigma_8$ values, to test whether the 3PCF can break the well-known degeneracy between the two parameters.}
   {We found that as a result of free-streaming, the strongest signal is found for quasi-isosceles and squeezed triangles; this signal increases for decreasing redshifts. 
    Among these configurations, elongated triangles, tracing the filamentary structure of the cosmic web, are the most affected by massive neutrinos, with a 3PCF signal increasing with $M_\nu$. 
    A complementary source of signal comes from right-angled triangles in $Q$. 
    Importantly, we found that the signatures of a $\sigma_8$ variation appear to be significantly different on elongated triangles in $\zeta$ and right-angled triangles in $Q$, suggesting that the 3PCF can be used to effectively break the $M_\nu - \sigma_8$ degeneracy. 
    These results open the possibility to use the 3PCF as a powerful complementary tool for constraining neutrino masses in current and future spectroscopic surveys such as DESI, Euclid, 4MOST, and the Nancy Grace Roman Space Telescope.}
   {}


   \maketitle

\section{Introduction}
\label{sec:introduction}

Neutrinos have nonzero mass, as first confirmed by the detection of their flavor oscillations \citep{Fukuda98}, providing clear evidence for physics beyond the standard model, where neutrinos are typically assumed to be massless. 
This has important implications for cosmology. 

According to the Big Bang paradigm, a thermal neutrino relic component, known as cosmic neutrino background, should exist, contributing to the total radiation energy density at early times, when still relativistic, and to the total matter density after the nonrelativistic transition \citep[for a comprehensive review]{Lesgourgues06}.
Since gravity is sensitive to the sum of neutrino masses $M_\nu \equiv \sum_i m_i$, cosmology is complementary to oscillation experiments, which instead measure the splitting of neutrino masses squared $\Delta m_i^2$ in determining the neutrino absolute mass scale, still one of the open problems of particle physics \citep[see][for a comprehensive review]{rev_part_phys}. 
Furthermore, reaching an accurate description of the imprint left by neutrinos on cosmological observables is essential to avoid systematics in the determination of cosmological parameters in current and upcoming spectroscopic surveys, such as the European Space Agency Euclid mission \citep{Euclid_definition}, the Dark Energy Spectroscopic Instrument \citep[DESI;][]{DESI_definition}, the 4-metre Multi-Object Spectroscopic Telescope \citep[4MOST;][]{4MOST_def}, and the Nancy Grace Roman Space Telescope \citep{nancy_grace}.

Due to their low masses, cosmic neutrinos have high thermal velocities even in the nonrelativistic regime; this has a significant effect on structure formation. 
Perturbations of the massive neutrino density field are indeed washed out on scales smaller than their free-streaming scale $\lfs$ \citep{Doroshkevich80, Bond80,  Lesgourgues06}, which is proportional to the average distance traveled by neutrinos during one Hubble time due to their thermal velocity. For a neutrino species of mass $m_i$, $\lfs$ evolves with redshift $z$ according to the relation \citep{Lesgourgues13}
\begin{equation}
    \label{eq:free-streaming}
    \lfs \approx 8.1 \, \frac{H_0 (1+z)}{H(z)} \, \frac{1 \, \ev}{m_i} \, \mpch \;,
\end{equation}
\noindent where $H(z)$ is the Hubble parameter at redshift $z$, and $H_0$ is its present-day value. 
On scales much larger than $\lfs$, neutrinos behave like cold dark matter (CDM).

Since perturbations of the cosmological fluid are the seeds for present-day observable structures, galaxy clustering represents an ideal probe for investigating the imprint of massive neutrinos on structure formation. 
In this framework, galaxies and the halos in which they reside are treated as tracers of the underlying matter field; the relation between their spatial distribution and the dark matter perturbations is known as the bias relation \citep{Kaiser84, Bardeen86, Desjacques16_bias_rev}. The evolution of perturbations and the bias relation were treated perturbatively by introducing linear and nonlinear terms in the dark matter perturbations \citep[][for a comprehensive review]{Bernardeau02}, with the latter becoming increasingly relevant on small scales.
Moreover, anisotropic effects relative to the line of sight (LOS), both linear and nonlinear, are introduced in the galaxy distribution by the peculiar velocities of galaxies, a phenomenon known as redshift-space distortions \citep[RSDs;][]{Kaiser87, Hamilton92, Fisher95, Scoccimarro99, Scoccimarro04, Taruya10}.

The statistical properties of the large-scale galaxy distribution are extracted by measuring the $N$-point statistics of the density field, starting with two-point statistics, specifically, the two-point correlation function (2PCF) in configuration space and its Fourier transform, the power spectrum, in Fourier space.
These statistics quantify the excess or deficit in the probability of finding pairs of galaxies with respect to a random distribution as a function of the distance between the two objects of the pair.

The effect of massive neutrinos on two-point statistics has been extensively studied in configuration and Fourier space. Neutrinos are found to suppress the total matter and CDM power spectra below the free-streaming scale \citep{HuEisenstein98, Brandbyge10, Viel10, Castorina15, VillaescusaNavarro18}, to affect RSDs by inducing a scale dependence in the linear growth rate $f$ and by modifying the root mean square of galaxy peculiar velocities  \citep{Marulli11, Verdiani25}, and to induce a scale-dependent bias even at large scales \citep{VillaescusaNavarro14, Castorina14}.
From $N$-body simulations, \citet{Castorina14} and \citet{Verdiani25} proved, respectively, that the linear bias depends solely on the variance of the CDM density field (the so-called universality in the CDM component), and RSDs in the linear regime are better described by assuming that the halo velocity field is unbiased with respect to the CDM velocity field alone.
The imprints of massive neutrinos were also studied through the reconstruction of cosmic microwave background (CMB) secondary anisotropies and their cross-correlation with CMB lensing and weak lensing signals \citep{Carbone16}, as well as through the cross-correlation of cosmic voids and CMB lensing \citep{Vielzeuf23}.
The effect of neutrinos has also been investigated at the scales of baryon acoustic oscillations \citep[BAO;][]{Peloso15, Parimbelli21}, recently focusing on systematics that may arise in neglecting neutrino masses in BAO reconstruction techniques \citep{NadalMatosas25}.

Cosmological analyses routinely use two-point statistics to constrain the sum of neutrino masses, often in combination with CMB data to break parameter degeneracies \citep[e.g.,][]{Sanchez14_confsp, Sanchez17_confsp, Grieb17, Ivanov20, Semenaite23, Moretti23_fullshape}.
The recent analyses from DESI \citep{DESI_fullshape, DESI_DRII, DESI_dr1_neutrinos}, combined with CMB data, yielded very stringent 95\% confidence level upper limits of $M_\nu \lesssim 0.07 \,\mbox{eV}$ while constraining $M_\nu \lesssim 0.4 \, \ev$ from two-point statistics alone.

However, two-point statistics provide a complete statistical description of the galaxy field only under the assumption of a perfectly Gaussian distribution  \citep{Bernardeau02}.
To quantify the non-Gaussian properties of the large-scale structure, higher-order statistics, such as the three-point correlation function \citep[3PCF;][]{Peebles80, FryGaztanaga93, FriemanGatzanaga94, Jing95, JingBoerner04} and its Fourier-space counterpart, the bispectrum \citep{Fry84, Scoccimarro99, Sefusatti06}, are needed.
Many sources of non-Gaussianity indeed act on the galaxy distribution, originating from nonlinearities involving the growth of perturbations \citep{Fry84}, RSDs \citep{Hivon95, Scoccimarro99}, and galaxy bias \citep{FryGaztanaga93, Fry94, FriemanGatzanaga94}, and potentially from several inflationary scenarios \citep{Verde2000, Celoria18, Meerburg19}. 
In particular, nonlinear effects, with their associated non-Gaussianity, are dominant on small scales. 
Since these scales are those on which neutrinos leave most of their characteristic signatures, statistics that quantify non-Gaussianity (i.e., higher-order ones) can be used to extract additional information with respect to lower-order ones.

Furthermore, two-point statistics are affected by degeneracies between parameters, in particular, by a strong degeneracy between $M_\nu$ and the cosmological parameter $\sigma_8$ \citep[e.g.,][]{Viel10, VillaescusaNavarro18}, the latter defined as the present-day standard deviation of the linear matter density field on a conventional scale of $8 \, \mpch$.
This degeneracy limits the possibility of obtaining precise constraints on $M_\nu$ from two-point statistics alone and requires exploring also higher-order statistics.

The first measurement of a bispectrum from $N$-body simulations including massive neutrinos was published in \citet{Ruggeri18}, quantifying the neutrino-induced suppression on the bispectrum and proving universality in the CDM component also beyond linear bias. 
The study of the halo and galaxy bispectrum from mock catalogs up to small scales proved its power in breaking degeneracies (including the $M_\nu - \sigma_8$ degeneracy) and tightening cosmological parameter constraints compared to the power spectrum alone \citep{Hahn20, Hahn21, KamalinejadSlepian25, KamalinejadSlepian26}.

Early studies of 3PCF only focused on specific configurations \citep{Gaztanaga05, McBride11, Marin13, Moresco14} or on the detection of acoustic features \citep{Gaztanaga09, deCarvalho20, Moresco21}, due to the high computational cost required by measures that rely on direct triplet counts. 
The introduction of an estimator based on spherical harmonic decomposition \citep[SHD;][]{SlepianEisenstein15, SlepianEisenstein18} has significantly reduced the computational cost, enabling more systematic analyses \citep{Slepian17_boss, Slepian17_boss_bao, Slepian18_boss}.
Another computational bottleneck concerns the modeling of the 3PCF, which is obtained by Fourier-transforming bispectrum models. 
While some perturbative approaches exploit one-dimensional fast Fourier transforms (FFT) at leading order \citep{SlepianEisenstein17_model, Sugiyama21}, more general methods require two-dimensional FFT \citep{Fang20} to perform the inversion \citep{Umeh21, Guidi23, Pugno25, Farina26}. 
The computational times of these approaches for sampling the parameter space properly are prohibitively long. This issue, however, can be overcome by developing emulators \citep{Guidi25_fullshape}.

In this evolving context, a systematic study of the effects of massive neutrinos on the 3PCF is still lacking. 
In addition to complementing bispectrum analyses, bridging this gap is crucial because configuration-space statistics are less sensitive to possible systematics arising from survey geometry, which instead introduce additional mode coupling in Fourier space that is far more challenging to account for in the modeling and estimators \citep{Philcox21, Pardede22}.  

We focus on the measurements of the halo 3PCF obtained from a large number of mock catalogs from $N$-body simulations that include a massive neutrino component.
This represents the first measurement of 3PCF in simulations implementing massive-neutrino cosmologies. 
In particular, we search for and quantify the signatures imprinted by massive neutrinos on the 3PCF, and we identify the structures that maximize the detectability of a potential neutrino signal by taking advantage of the power of the 3PCF to infer clustering as a function of the triangle scale and shapes. 
Moreover, we exploit this capability to disentangle the effect of massive neutrinos from variations in $\sigma_8$.

This paper is organized as follows. 
In Sect. \ref{sec:methods_data} we provide an overview of the methods and data employed in this analysis, defining the adopted statistics (\ref{subs:definitions}) and describing the set of simulations used (\ref{subs:simulations}), the estimators considered (\ref{subs:estimates}), the dataset produced from the measurements together with the estimation of covariance (\ref{subs:dataset_covariance}), and the framework developed for the neutrino detectability analysis (\ref{subs:det_metrics}).
In Sect. \ref{sec:results} we present our results, focusing on the triangle scale (\ref{subs:singlescale_analysis} and \ref{subs:scale_analysis}) and shape (\ref{subs:shape_analysis}) dependence of the signal from massive neutrinos, and the possibility of breaking the $M_\nu - \sigma_8$ degeneracy with the 3PCF (\ref{subs:degeneracy}).
Finally, in Sect. \ref{sec:conclusions} we draw our conclusions.

\section{Methods and data}
\label{sec:methods_data}

\subsection{Clustering statistics}
\label{subs:definitions}

The probability $\dif P$ of finding a triplet of objects inside the comoving volumes $\dif V_1$, $\dif V_2$, and $\dif V_3$, separated by the comoving distances $s_{12}$, $s_{13}$, and $s_{23}$, can be written as
\begin{equation}
    \label{eq:zeta_definition}
    \begin{aligned}
        \dif P = \bar{n}^3 \, [ 1 + \xi(s_{12}) + \xi(s_{13}) + \xi(s_{23}) + \zeta(&s_{12}, s_{13}, s_{23})] \\
        &\dif V_1\dif V_2\dif V_3\; ,
    \end{aligned}
\end{equation}
\noindent where $\bar{n}$ is the average number density of objects, and $\xi$ and $\zeta$, are the 2PCF and connected 3PCF, respectively \citep{Peebles80}.
Unlike the 2PCF, which only encodes scale information (being exclusively dependent on the separation between pairs of objects), the 3PCF is the lowest-order clustering statistics able to also provide information about the shape of structures (since size and shape both characterize triangles).

In redshift space, $\xi$ and $\zeta$ also depend on the orientation between a given pair or triplet, respectively, and the LOS unit vector $\vec{\hat{n}}$ due to the action of RSDs, which break the assumption of isotropy by introducing the privileged direction defined by $\vec{\hat{n}}$. 
This splits clustering statistics into an isotropic component, defined by averaging them in redshift space over all the possible LOS directions, which solely depends on pair/triplet separations, and an anisotropic component that retains the specific directional dependence.
We only focus on the isotropic component of the statistics we considered here and aim to extend it with anisotropic information in a future work.  

We complemented the information provided by the isotropic 3PCF by also considering the reduced 3PCF $Q$ \citep{GrothPeebles77}, defined as 
\begin{equation}
   \label{eq:q_definition}
           Q(s_{12}, s_{13}, s_{23}) = \frac{\zeta(s_{12}, s_{13}, s_{23})}{\xi_0(s_{12})\xi_0(s_{13}) + \xi_0(s_{13})\xi_0(s_{23})+ \xi_0(s_{23})\xi_0(s_{12})}\; ,
\end{equation}
\noindent where $\xi_0$ denotes the monopole of the 2PCF (i.e., the isotropic component of the 2PCF).
The reduced 3PCF provides a natural combination of $\zeta$ and $\xi$; since it can be demonstrated that in hierarchical scenarios, $\zeta \propto \xi^2$ to a good approximation  \citep{PeeblesGroth75}, this quantity is on the order of unity on all scales by definition, and it is explicitly independent of $\sigma_8$ by construction \citep[see Eq. 6 in][]{Moresco21}. 

\subsection{Estimators}
\label{subs:estimates}

We estimated the 2PCF with the natural estimator \citep{Peebles73}
\begin{equation}
    \label{eq:natural_estimator}
    \hat{\xi}(s, \mu) = \frac{DD(s,\mu)}{RR(s,\mu)} -1\; ,
\end{equation}
\noindent where $\mu$ is the cosine of the angle between the pair and the LOS direction, and $DD$ and $RR$ are the pair counts in the data and in a random distribution of unclustered objects with the same geometry as the data catalog, respectively.
In our case, that is, a simulation box with periodic boundary conditions (as detailed in Sect. \ref{subs:simulations}), this estimator is equivalent to the usual Landy-Szalay estimator \citep{Landy93}. 
Moreover, periodicity allowed us to compute the $RR$ term analytically.
We then obtained the 2PCF monopole by numerically averaging $\hat{\xi}$ over $\mu$.

We estimated the isotropic connected 3PCF with the SHD estimator introduced in \citet{SlepianEisenstein15}, which has the advantage of scaling with the number of objects $N$ as $\mathcal{O}(N^2)$, rather than $\mathcal{O}(N^3)$, as in the case of previous estimators relying on direct triplet counts. 
This approach is based on parameterizing a given triangle of sides $s_{12}, s_{13}$, and $s_{23}$ with two of its sides, for example, $s_{12}$ and $s_{13}$, and the angle $\theta$ between them. The third side $s_{23}$ is then reobtained as a function of $s_{12}$, $s_{13}$, and $\theta$.
This parameterization allowed us to expand the dependence of the isotropic 3PCF $\zeta(s_{12}, s_{13}, \theta)$ on $\theta$ into Legendre polynomials \citep{Szapudi04}, with coefficients given by the corresponding Legendre multipoles $\zeta_\ell(s_{12}, s_{13})$ \citep[see Eq. 11 in][]{Guidi25_fullshape}.
Therefore, the full estimator $\hat{\zeta}(s_{12}, s_{13}, \theta)$ for the connected 3PCF can be written as a function of an estimator $\hat{\zeta}_{\ell}(s_{12}, s_{13})$ for the isotropic Legendre multipoles as 
\begin{equation}
    \label{eq:full_3pcf_estimator}
    \hat{\zeta}(s_{12}, s_{13}, \theta) = \sum_{\ell = 0}^{\ell_{\max}} \hat{\zeta}_{\ell}(s_{12}, s_{13}) \, \leg_\ell(\cos\theta)\; ,
\end{equation}
\noindent where $\ell_{\max}$ is the highest-order multipole included in the expansion.
The isotropic multipoles are estimated as 
\begin{equation}
    \label{eq:zeta_ell_estimator}
    \hat{\zeta}_{\ell}(s_{12}, s_{13}) = \frac{DDD_\ell - 3DDR_\ell + 3DRR_\ell - RRR_\ell}{RRR_0}\; ,
\end{equation}
\noindent where $DDD_\ell$, $DDR_\ell$, $DRR_\ell$, and $RRR_\ell$ are the multipoles of the Legendre expansion of the data-data-data, data-data-random, data-random-random, and random-random-random triplet counts, respectively. 
This expression is analogous to the traditional \citet{SzapudiSzalay98} direct triplet count estimator, applied to the case of the 3PCF multipoles.
The evaluation of the terms in Eq. \ref{eq:zeta_ell_estimator} was detailed in \citet{SlepianEisenstein15}.
Again, the periodicity of the simulation box allows for the analytical computation of the monopole of the random counts $RRR_0$.

The efficiency of the SHD estimator is reduced for nearly isosceles triangle configurations ($s_{12} \simeq s_{13}$), since a much larger number of multipoles ($\ell_{\max} > 30$) are needed to properly reconstruct the shape of $\zeta$ when $\theta \to 0$ \citep[e.g.,][]{Veropalumbo21}.
For this reason, we adopted the quantity introduced by \citet{Veropalumbo22},
\begin{equation}
    \label{eq:eta_definition}
    \eta \equiv \frac{|s_{13} - s_{12}|}{\Delta s}\; ,
\end{equation}
\noindent which can be used to exclude, by setting $\eta > \eta_{\min}$, triangles that progressively deviate from the isosceles configuration \citep[see also][]{Guidi23, Guidi25_fullshape, Farina26}.
For our measurements, we used the implementation of the estimators in Eqs. \ref{eq:natural_estimator} and \ref{eq:full_3pcf_estimator} provided in the publicly available software \texttt{MeasCorr}\footnote{\url{https://gitlab.com/veropalumbo.alfonso/meascorr}} \citep{Farina26}.

\subsection{Simulation dataset}
\label{subs:simulations}

We used the \textsc{Quijote}\footnote{\url{https://quijote-simulations.readthedocs.io/en/latest/types.html}} suite of $N$-body simulations \citep{quijote_paper}, which provides a large number of realizations to assess the effect on several statistics of variations in the cosmological parameters and to estimate covariance matrices. 
They were run using the tree particle mesh-smoothed particle hydrodynamics code \textsc{Gadget-III} \citep{Springel05}.
The suite contains massless- and massive-neutrino simulations, whose main properties we describe below. 

The fiducial cosmology of the simulations corresponds to a flat $\Lambda$CDM Universe with cosmological parameters in agreement with the latest constraints by Planck \citep{Planck18}. In particular, it is characterized by the sum of neutrino masses $M_\nu = 0 \, \ev$ and $\sigma_8 = 0.834$. 
Massive-neutrino simulations assume three degenerate neutrino masses and were implemented by using the particle-based method \citep{Brandbyge08, Viel10}, in which neutrinos are described as a collisionless and pressureless fluid discretized into particles. The simulations we considered followed the evolution of $512^3$ CDM particles plus, if $M_\nu \neq 0$, $512^3$ neutrino particles, in a periodic cubic box of side length $L = 1$ \gpch, and have a softening length of 50 \kpch. The initial conditions (ICs) of the simulations were generated at $z_i = 127$. 
Displacements and peculiar velocities of particles were computed either with the Zeldovich approximation \citep[ZA;][]{Zeldovich70} or with the second-order Lagrangian perturbation theory \citep[2LPT;][]{Bernardeau02} in massive and massless neutrino models, respectively.
 In addition to peculiar velocities, neutrino particles were assigned thermal velocities randomly drawn from a Fermi-Dirac distribution at $z_i$.
Halos were identified by running the friend-of-friends algorithm \citep{Davis85} with a linking length parameter $b = 0.2$ on the CDM particles.
Only halos containing at least 20 CDM particles were saved, corresponding to a minimum halo mass $M_{\min} \approx 1.3 \times 10^{13} \, h^{-1} \, M_{\sun}$.

We considered massive neutrino simulations that were run for three different values of $M_\nu = 0.1, \, 0.2$, and $0.4 \, \mbox{eV}$ (with the remaining cosmological parameters fixed at their fiducial values), labeled \mnup, \mnupp, and \mnuppp\, respectively (500 realizations per simulation).
The $z = 0$ values of the free-streaming scale in these cosmologies according to Eq. \ref{eq:free-streaming} are $\lfs \sim 240, 120$, and $60 \, \mpch$ for $M_\nu = 0.1, 0.2$, and $0.4 \, \ev$, respectively (where we assumed $m_i = M_\nu/3$), with only a $\sim 10 \%$ variation in the redshift range $0 \leq z \leq 2$ relevant for this work assuming the \textsc{Quijote} cosmology.

To study the degeneracy between $M_ \nu$ and $\sigma_8$, we complemented this set with massless-neutrino simulations differing from the fiducial cosmology only in the value of $\sigma_8$, with $\sigma_8 = 0.849$, and $\sigma_8 = 0.819$, labeled \sigmap\ and \sigmam, respectively (500 realizations per simulation). The control sample of the massive-neutrino and varying-$\sigma_8$ simulations was made of two sets of 500 realizations each of the fiducial cosmology, run with ZA and 2LPT ICs, and labeled \fid\ and \fidza.
We also included $2\,000$ additional realizations of the fiducial cosmology run with 2LPT ICs for the numerical estimation of covariance. 
For all the realizations, we moved halos to redshift space by computing RSDs along the $\vec{\hat{z}}$ axis of the simulations.

\begin{figure*}
\centering
   \includegraphics[width=17cm]{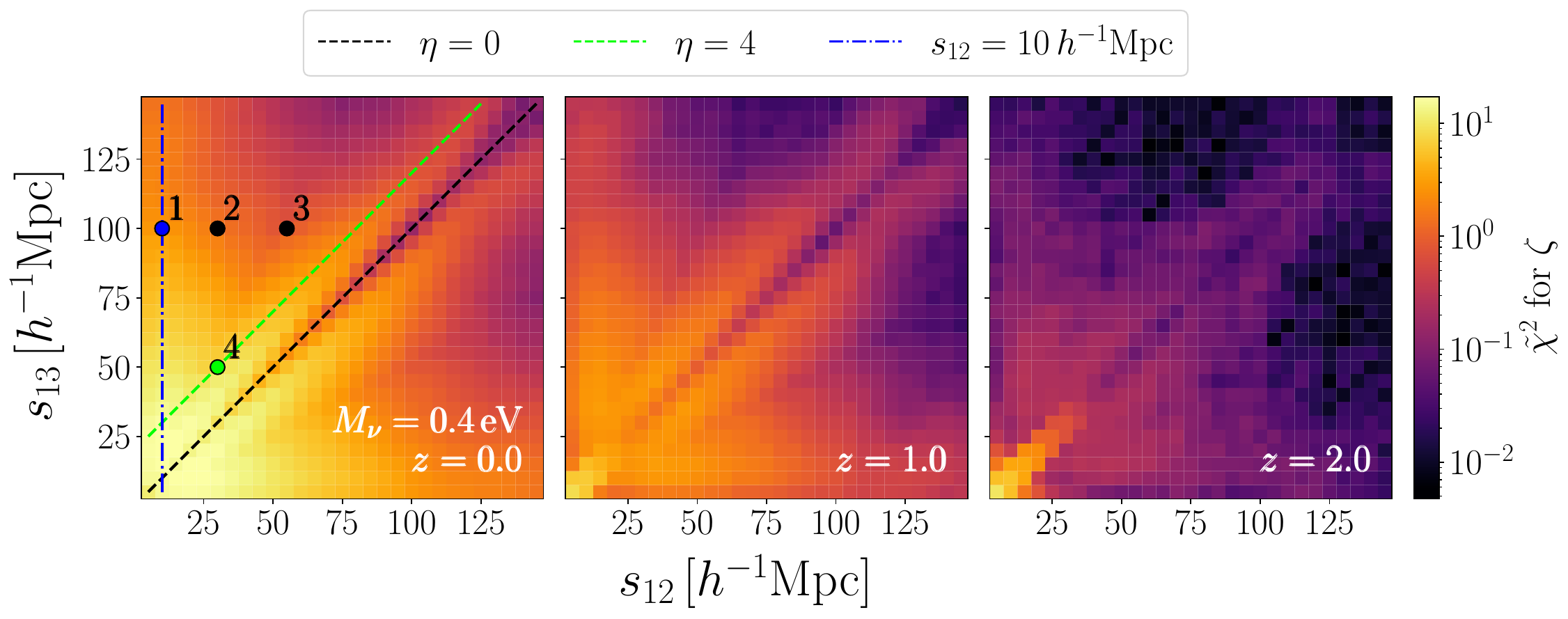}
     \caption{Values of the parameter $\redchiq(s_{12}, s_{13})$ defined in Eq. \ref{eq:single-scale_chi_det} for the single-scale connected 3PCF obtained for $M_\nu = 0.4$ \ev.
     Each panel corresponds to a different redshift. From left to right, $z = 0, 1$, and 2.
     The lines overplotted on the left panel are taken as representative of regions with a stronger signal, and they identify isosceles triangles (dashed black line with $\eta = 0$, where the signal is only enhanced on scales $\lesssim 30$ \mpch), quasi-isosceles triangles with $\eta = 4$ (dashed green line), and triangles with $s_{12} = 10$ \mpch\ (dash-dotted blue line).
     The numbered circles on the lines identify some ($s_{12}, s_{13}$) configurations that correspond in increasing order from 1 to 4 to (10,100), (30,100), (55,100), and (30,50) \mpch. For them, we plot the single-scale $\zeta$ in Fig. \ref{fig:single_scales_zeta}.}
     \label{fig:single_scale_chi2_zeta}
\end{figure*}

\subsection{Measurements and covariance}
\label{subs:dataset_covariance}

We measured the 2PCF monopole for the halo catalogs of the simulations listed in Sect. \ref{subs:simulations} at $z = 0,1,2$. 
For these redshifts, the number of halos identified in each realization is $\sim 4 \times 10^5$, $2 \times 10^5$, and $\sim 4.4 \times 10^4$.
We chose separations ranging from $s_{\min} = 1$ \mpch\ to $s_{\max} = 150$ \mpch\ and linearly spaced bins of width $\Delta s = 1$ \mpch\ and $\Delta \mu = 0.01$.
For the connected 3PCF, we included all triangles with side lengths from $s_{\min} = 2.5$ \mpch\ to $s_{\max} = 147.5$ \mpch, considering a bin width $\Delta s = 5$ \mpch, and we estimated the isotropic multipoles $\zeta_\ell$ up to $\ell_{\max} = 10$, which for the vast majority of triangle configurations represents an optimal balance between computational cost and information content. 

For each set of simulations, we averaged the measured multipoles over the different realizations and followed two approaches.
In the first approach, which we refer to as the single-scale approach, we fixed two triangle sides $s_{12}$ and $s_{13}$ and computed $\zeta(s_{12}, s_{13}, \theta)$ and $Q(s_{12}, s_{13}, \theta)$. We obtained the average $\hat{\zeta}$ for each set of simulations by substituting the average multipoles in Eq. \ref{eq:full_3pcf_estimator}, and we used it together with the average 2PCF monopole in Eq. \ref{eq:q_definition} to obtain the average $\hat{Q}$.
In the second approach, we computed the average  $\hat{\zeta}$ and $\hat{Q}$ of each simulation set for all the possible side-binned triangles obtained by adopting the ordering $s_{12} \leq s_{13} \leq s_{23}$. We refer to this scheme as the all-scales approach.

To evaluate the denominator of $Q$, we linearly interpolated the average 2PCF monopole at the values of $s_{12}, s_{13}$, and $s_{23}$ in Eq. \ref{eq:q_definition} in both approaches.   
Moreover, since the 2PCF monopole changes sign for $s \sim 120$ \mpch, the denominator of $Q$ can exhibit zero crossings when at least one triangle side is above this scale. 
However, the separation $s$ at which $\xi_0$ changes sign depends on the realization. We therefore inserted already averaged quantities in Eq. \ref{eq:q_definition} instead of averaging after estimating $Q$ for each realization. 
As an additional precaution, we restricted the analysis of the reduced 3PCF to configurations for which all sides are smaller than 110 \mpch, in which case the average monopole of the 2PCF remains positive.

We numerically estimated the covariance matrix of the multipoles of the 2PCF and 3PCF from the $2\,000$ \fid\ mocks. 
The 3PCF multipole covariance depends on two triangle side pairs $\vec{p} = (s_{12}, s_{13})$, $\vec{p'} = (s_{12}', s_{13}')$ and two multipole indexes $\ell, \ell'$, so we denoted it with $\hat{C}_{\zeta, \ell \ell'}(\vec{p}, \vec{p'})$. 
The single-scale and all-scales covariance matrices can then be obtained as
\begin{align}
    \label{eq:cov_single_scale_zeta}
    \hat{C}_{\zeta, ij}(\vec{p}) &= 
    \sum_{\ell, \ell' = 0}^{\ell_{\rm{max}}} \hat{C}_{\zeta, \ell \ell'}(\vec{p}, \vec{p}) \, \leg_\ell(\cos\theta_i) \, \leg_{\ell'}(\cos \theta_j)\; ,\\
    \label{eq:cov_all_scale_zeta}
    \hat{C}_{\zeta, \vec{t}\vec{t'}} &= \sum_{\ell, \ell' = 0}^{\ell_{\rm{max}}} \hat{C}_{\zeta, \ell \ell'}(\vec{p}, \vec{p'}) \, \leg_\ell(\vec{t}) \, \leg_{\ell'}(\vec{t'})\; ,
\end{align}
\noindent with $\vec{t} = (s_{12}, s_{13}, s_{23})$, $\vec{t'} = (s_{12}', s_{13}', s_{23}')$. 
We rescaled all covariances in a volume of 10 \gpchc, taken as an ideal representative of a redshift bin of a stage-IV survey \citep{Albrecht06} at $z \sim 1$, such as for the Euclid Wide Survey \citep{Euclid_wide_survey}.
We estimated that even extending it to lower or higher redshifts does not significantly affect our findings.
All the measurements of 3PCF obtained for this analysis are available among the data products of the \textsc{Quijote} suite.\footnote{\url{https://quijote-simulations.readthedocs.io/en/latest/3PCF.html}}

\subsection{Detectability metrics}
\label{subs:det_metrics}

The simplest parameter we defined to quantify the neutrino signal is an error-weighted difference, which we simply refer to as detectability, 
\begin{equation}
    \label{eq:def_detectability}
    \mathrm{DET}_i \equiv \frac{\hat{f}_i(M_\nu) - \hat{f}_i(M_\nu = 0)}{\sqrt{2}\, \sigma_i}\; ,
\end{equation}
\noindent where the numerator is the difference between the averages of a given statistics (e.g., the average $\hat{\zeta}$ or $\hat{Q}$), estimated in one of the massive-neutrino cosmologies and in the fiducial cosmology from the \fidza\ mocks. 
This parameter provides an estimate of the detectability in a specific configuration identified by the index $i$, denoting a generic bin in which the difference is evaluated. The error is obtained as $\sigma_i = \sqrt{\hat{C}_{f,ii}}$, where $\hat{C}_f$ is the covariance matrix of $f$. 

We also generalized the element-wise detectability in Eq. \ref{eq:def_detectability} by introducing a metric that estimates the detectability over a given range of configurations, accounting for their correlation, defined by the parameter 
\begin{equation}
    \label{eq:single-scale_chi_det}
    \redchiq(\vec{p}, M_\nu) \equiv \frac{1}{N_\theta} \, \sum_{i,j=1}^{N_\theta} 
    \Delta \hat{f}(\theta_i; \,\vec{p}, M_\nu) \, \hat{C}^{-1}_{f,ij} (\vec{p})\, \Delta \hat{f}(\theta_j; \, \vec{p}, M_\nu)\; ,
\end{equation}
\noindent where $i$ and $j$ run on the angles formed by $s_{12}$ and $s_{13}$, and the differences $\Delta \hat{f}$, with $\hat{f} = \hat{\zeta}$ or $\hat{Q}$, are defined in the same way as in the numerator of Eq. \ref{eq:def_detectability}.
We defined this parameter $\redchiq$ in analogy with the standard definition of the reduced chi-squared, except that in place of the model, we inserted the estimated fiducial statistic.

This parameter also allowed us to provide a quantitative assessment of the statistical significance of the signal from neutrinos of total mass $M_\nu$.
In particular, it is possible to compute the $p$ value associated with a given value of $\redchiq(\vec{p}, M_\nu)$, that is, the tail integral of a chi-squared distribution $\mathcal{F}(\chiq; N_\theta)$ with $N_\theta$ degrees of freedom, 
\begin{equation}
    \label{eq:p-value}
    p \equiv \int_{N_\theta \: \redchiq}^\infty \dif\chiq \, \mathcal{F}(\chiq; N_\theta)\; ,
\end{equation}
\noindent and then convert it into an equivalent Gaussian statistical significance $Z\sigma$, where $Z$ is implicitly defined as follows \citep{Cowan11}:
\begin{equation}
    \label{eq:signif}
    2\int_Z^\infty \dif z \, \mathcal{N}(z) = p\; .
\end{equation}
\noindent Here, $\mathcal{N}$ is the standard Gaussian distribution, characterized by zero mean and unitary variance. 

To account for the fact that the precision matrix obtained by inverting a numerically estimated covariance matrix is a biased estimator of the parent population precision matrix, we applied the correction factor prescribed by \citet{Hartlap07}.
This consists of multiplying each inverse covariance matrix by the factor $(N_m - n_d - 2)/(N_m - 1)$, where $N_m$ is the number of mocks used to estimate the covariance, and $n_d$ is the dimensionality of the data vector. 

As a final note, we emphasize that although the various indicators defined may appear quite different, they are all based on the same underlying principle,  quantifying the deviation of a given measurement from the fiducial one, weighted by the associated uncertainty.
However, depending on the case, it is useful to use one or the other, as they provide complementary information.

\section{Results}
\label{sec:results}
We estimated the detectability of massive neutrinos in different configurations. Below, we present our main results, starting with the 3PCF analysis in single-scale configurations and then moving to the all-scales approach.

\subsection{Single-scale analysis}
\label{subs:singlescale_analysis}

\begin{figure*}
\centering
   \includegraphics[width=17cm]{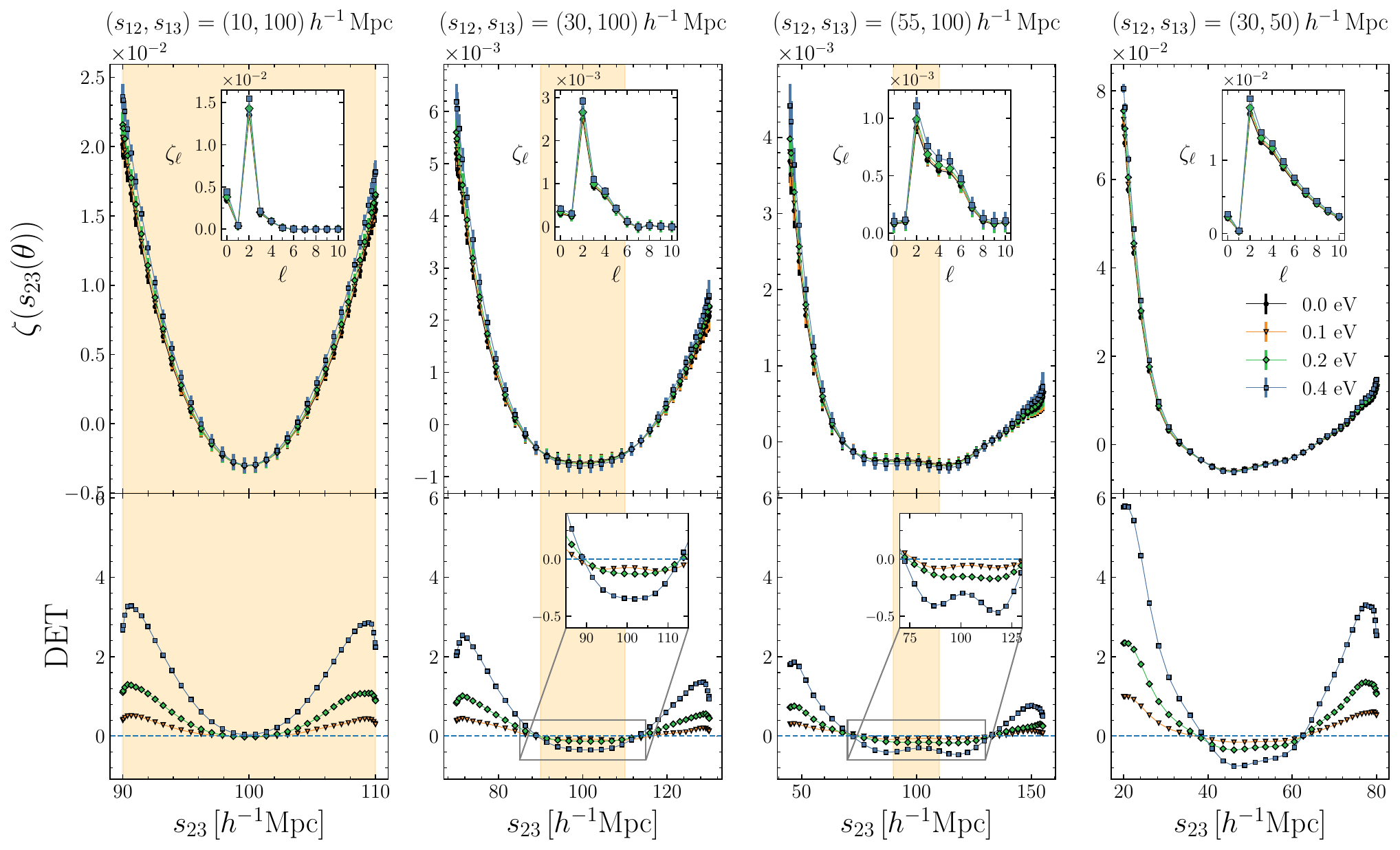}
     \caption{Single-scale connected 3PCF for the triangle configurations selected in Fig. \ref{fig:single_scale_chi2_zeta} (upper plots of each panel), as indicated in the top label. 
     We show the results at $z = 0$, with a different color for each neutrino mass, as indicated in the legend.
     The inset plots in the upper row show the multipoles $\zeta_\ell(s_{12}, s_{13})$ from which the 3PCF was reconstructed.
     The lower plots show the corresponding detectabilities as a function of $s_{23}$ (Eq. \ref{eq:def_detectability}).
     The dashed blue line marks the zero detectability level.
     The orange shaded areas show the region $90 \, \mpch \leq s_{23} \leq 110 \, \mpch$, corresponding to the expected location of the BAO peak. 
     Where the BAO scales do not cover the full $s_{23}$ range, we include a zoom-in on the detectability in that region to better visualize the effect of neutrinos in those ranges.}
     \label{fig:single_scales_zeta}
\end{figure*}

We computed the parameter $\redchiq$ defined in Eq. \ref{eq:single-scale_chi_det} for all the values of $M_\nu$ and redshift.
The covariance matrix (Eq. \ref{eq:cov_single_scale_zeta}) is singular if $N_\theta > \ell_{\max} + 1$. 
For this reason, we chose to estimate $\zeta$ in ten evenly spaced angular bins, with $0 \leq \theta \leq \pi$.
For $Q$, we excluded the first bin, corresponding to $\theta = 0$. This ensured that the third side $s_{23}$ was not smaller than the minimum separation at which we measured the 2PCF monopole, required for its interpolation.

We show the results for $\zeta$ in Fig. \ref{fig:single_scale_chi2_zeta} for the illustrative case $M_\nu = 0.4$ \ev at all redshifts.
As a general trend, we find that the detectability of neutrinos increases as the redshift decreases.
At a given redshift, it appears mostly concentrated in specific triangle configurations, identified by the brightest colors in the figure.
A first set of configurations is defined by the isosceles ones (i.e., the diagonal of the matrix) at small scales, up to $s_{12} = s_{13} \sim 30$ \mpch. 
This is particularly evident considering the results for $z = 2$. 
Conversely, the signal drops on larger scales, due to an increase in the variance of the multipoles used to reconstruct the angular dependence of $\zeta$, as detailed in Appendix \ref{app:multipoles}. 
Other configurations that show high values of $\redchiq$ are quasi-isosceles ($s_{12} \approx s_{13}$) and squeezed triangles ($s_{13} \gg s_{12}$).
These configurations correspond to the two regions close to the diagonal, and to the areas adjacent to the left and lower sides of the figure, respectively, except for the bottom left corner (where $s_{12}$ and $s_{13}$ are both small) in the squeezed case.

The behavior in the detectability of the neutrino signal in the 3PCF can be explained as a consequence of free-streaming.
For a fixed amplitude of the primordial density fluctuations $A_\mathrm{s}$, high neutrino masses produce lower values of $\sigma_8$, since the suppression of the matter clustering below $\lfs$ is stronger. 
However, the massive-neutrino simulations of the \textsc{Quijote} suite were created with fixed $\sigma_8 = 0.834$, hence with a larger $A_\mathrm{s}$ for increasing $M_\nu$ (in detail, $10^9 A_\mathrm{s} =  2.13, 2.25, 2.40$, and $2.74$ for $M_\nu = 0.0, 0.1, 0.2$, and $0.4$ \ev, respectively). 
As a leading effect, the different values of $A_\mathrm{s}$ result in an overall rescaling of all clustering statistics, with scaling factor depending on $M_\nu$.
This implies that the differences between the 3PCF measured in the massive- and massless-neutrino cases are larger, therefore increasing the detectability, when the 3PCF takes higher values. 
This occurs on small scales, while on larger scales, where nonlinear effects become negligible, the 3PCF values drop rapidly.
This picture is consistent with the overall increase in the $\redchiq$ values as $z$ decreases, and with the configurations that maximize it at fixed redshift, since isosceles, quasi-isosceles, and squeezed triangles are precisely those for which at least one of the three sides can probe nonlinear or mildly nonlinear scales. 
As further confirmation, moving toward the bottom-left corner of the figures, the values of $\redchiq$ increase, reflecting the fact that in this case, all three sides become progressively smaller.

The results obtained for the reduced 3PCF are shown in Fig. \ref{fig:single_scale_chi2_q}.
With respect to the connected 3PCF, the neutrino detectability for the reduced 3PCF is lower overall because the errors from the propagation of the uncertainties on $\zeta$ and $\xi$ are larger. 
At all redshifts, the highest values of $\redchiq$ are localized in isosceles configurations and increase moving from large to small scales.

\begin{figure*}
\centering
   \includegraphics[width=17cm]{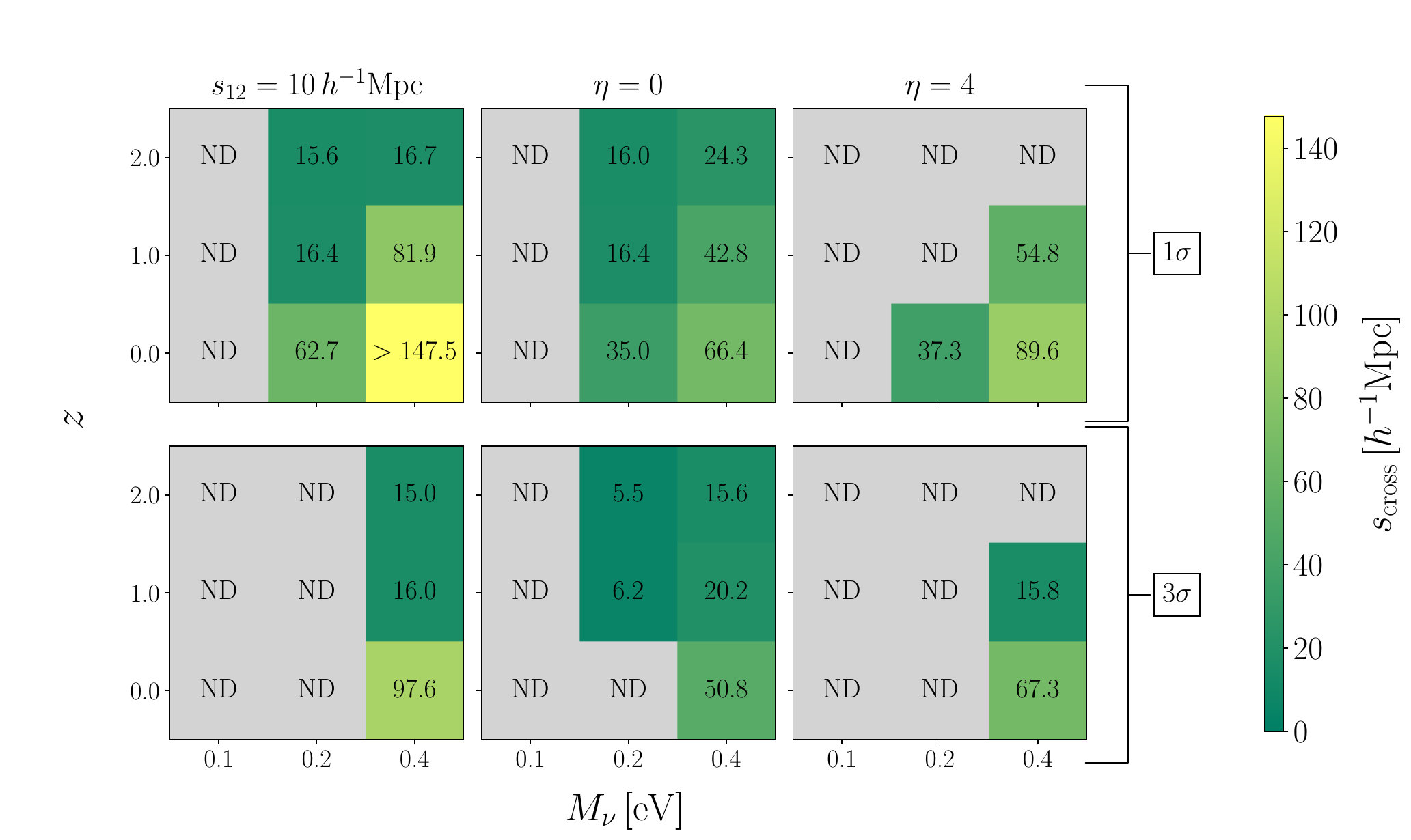}
     \caption{Detectability matrices of massive neutrinos. 
     Each matrix element shows the scale $s_{\mathrm{cross}}$ (Eq. \ref{eq:scale_crossing}) below which, considering all triangles with a scale larger than $s_{\mathrm{cross}}$, we obtained a significant detection of the signal from massive neutrinos in the connected 3PCF. 
     The matrices in the upper and lower rows show the values for a $1\sigma$ and $3\sigma$ statistical significance (Eq. \ref{eq:signif}), respectively, computed for a volume of $10 \, \gpchc$.
     In each matrix, $M_\nu$ increases from left to right, and redshifts increase from bottom to top.
     We show the results for the three configurations corresponding to the lines in Fig. \ref{fig:single_scale_chi2_zeta}, i.e., from left to right: triangles with $s_{12} = 10 \, \mpch$, isosceles triangles ($\eta = 0$), and quasi-isosceles triangles with $\eta = 4$.
     The quantity $s_{\mathrm{cross}}$ represents one of the two sides, $s_{12}$ or $s_{13}$, depending on the configuration: in particular, in Eq. \ref{eq:scale_crossing} we set $s = s_{13}$ for triangles with fixed $s_{12}$, and $s = s_{12}$ for isosceles and  $\eta = 4$ triangles. 
     Brighter colors represent better detection levels, i.e., occurring at larger scales.
     The label ND stands for ``not detectable'' above the specified significance threshold at any scale.
     A lower limit is indicated whenever the signal is detectable above a given threshold of statistical significance over the entire range of scales considered in our analysis.
     }
     \label{fig:summary_zeta}
\end{figure*}

In Fig. \ref{fig:single_scale_chi2_zeta} we identify some sets of configurations from the regions in which the signal is maximized, corresponding to the lines shown in the leftmost panel, with equations $\eta = 0$, $\eta = 4$, and the vertical line $s_{12} = 10$ \mpch.
To visually inspect the effect of neutrinos directly on the 3PCF, we chose some specific configurations, on and off the identified lines, for comparison, corresponding to $(s_{12}, s_{13}) = (10,100), (30,100), (55,100)$, and $(30,50) \, \mpch$; these are shown in the figure with the numbered markers. 
For these, we show in Fig. \ref{fig:single_scales_zeta} the single-scale connected 3PCF at $z = 0$ for all the neutrino masses $M_\nu$ as a function of the third side $s_{23}$ and their detectabilities.
Additionally, we show the measured multipoles $\zeta_\ell(s_{12}, s_{13})$ from which the functions were estimated.
We avoid showing $\zeta$ for $(s_{12}, s_{13})$ with $\eta = 0$ because the reconstruction provided by the SHD estimator is, as expected, not accurate. 
In particular, $\zeta$ is characterized by a strong steepening for $s_{23} \to 0$ (or equivalently, for $\theta \to 0$), which cannot be accurately reconstructed by $\ell_{\max} = 10$ \citep[e.g., see Fig. 2 in][]{Veropalumbo21}.
The three leftmost configurations in Fig. \ref{fig:single_scales_zeta} show an optimal reconstruction for $\ell_{\max} = 10$, since the signal encoded by the highest multipoles is negligible with respect to the lowest ones. 
For $(s_{12}, s_{13}) = (30,50)$ \mpch\ (rightmost panel), the first ten multipoles still provide a good reconstruction of $\zeta$, although a small amount of residual signal remains confined to $\ell > 10$.

From the detectability plots, it is evident that the signal is captured mainly by configurations that minimize or maximize $s_{23}$, corresponding to the angles $\theta = 0$ and $\theta = \pi$.
This implies that the signal is mostly driven by elongated triangles, tracing the filamentary structure of the cosmic web. 
For intermediate values of $s_{23}$, each mass is either not detectable (as in the case $s_{12} = 10 \, \mpch, s_{13} = 100$ \mpch, where $\detec \approx 0$ for $s_{23} \approx 100$ \mpch) or only poorly detectable, with negative detectability values, meaning that $\zeta(M_\nu) < \zeta(M_\nu = 0)$, as in the case of $(s_{12}, s_{13}) = (30,50)$ \mpch. 

An analogous plot for the single-scale reduced 3PCF is shown in Fig. \ref{fig:single_scale_q}.
Although the neutrino signal is, as previously noted, smaller in amplitude, the behavior of $Q$ as a function of $M_\nu$ differs substantially from that of $\zeta$.
In particular, $Q$ becomes flatter as $M_\nu$ increases.
This produces a different detectability profile as a function of the third side, with negative values for filamentary configurations and positive values for intermediate, more rounded configurations.
The same profiles show that the signal is equally driven by filamentary and intermediate configurations, complementary to the behavior obtained with $\zeta$, where the signal is instead largely dominated by the contribution of filamentary structures.

We also inspected the specific contribution of the signal of massive neutrinos at BAO scales. 
The single-scale 3PCF exhibits a small peak at $s_{23} \sim 100$ \mpch\ when, for increasing $\theta$, the two fixed sides $s_{12}$ and $s_{13}$ allow the third side $s_{23}$ to cross the BAO scales \citep{Gaztanaga09}. 
This peak has a nontrivial interplay in defining the shape of the 3PCF, possibly canceling out with the 3PCF dip, giving $\zeta$ a flat shape for $\theta \sim \pi/2$, or resulting in a small peak embedded in the dip, depending on the considered configuration. 
The behavior is related to the angular spreading of the feature. A smaller angular spreading indeed implies a higher visibility of the peak \citep{Moresco21}. Equivalently, the visibility of the peak is higher when BAO scales are mapped in a smaller range of $s_{23}$.

We explicitly chose the configurations $(s_{12}, s_{13}) = (10, 100), (30, 100)$, and $(55,100)$ \mpch, so that the BAO feature was progressively concentrated in a smaller $s_{23}$ range, highlighted in light orange in Fig. \ref{fig:single_scales_zeta}.
For $(s_{12}, s_{13}) =(10, 100)$ \mpch, the area spans the entire $s_{23}$ interval, so no distinctive feature is observable. 
For $(s_{12}, s_{13}) = (30, 100)$ \mpch, $\zeta$ is flatter at the minimum with respect to the previous case, without showing any local maximum, and an extremely shallow peak is visible in the $M_\nu = 0.1$ \ev\ detectabilities at $\sim 100$ \mpch.
Finally, for $(s_{12}, s_{13}) =(55, 100)$ \mpch, a small peak is visible in $\zeta$ for all the values of $M_\nu$ and in the detectabilities of $M_\nu = 0.4$ and 0.1 \ev.
However, we cannot derive any statistically significant effect of the dependence of the amplitude of the BAO feature on $M_\nu$ because the uncertainties on $\zeta$ in the BAO region are too large. 
We note that even increasing the effective volume to $V_{\mathrm{eff}} = 500$ \gpchc does not allow a statistically significant detection at those scales.

In $Q$, we do not identify any distinctive BAO feature. This is due to the adopted scale cut discussed in Sect. \ref{subs:dataset_covariance}, which spreads any possible BAO signal over a large portion of the $s_{23}$ range (as in the case of the leftmost panel in Fig. \ref{fig:single_scales_zeta}).

\subsection{Scale dependence of the neutrino signal}
\label{subs:scale_analysis}

\begin{figure*}
\centering
   \includegraphics[width=17cm]{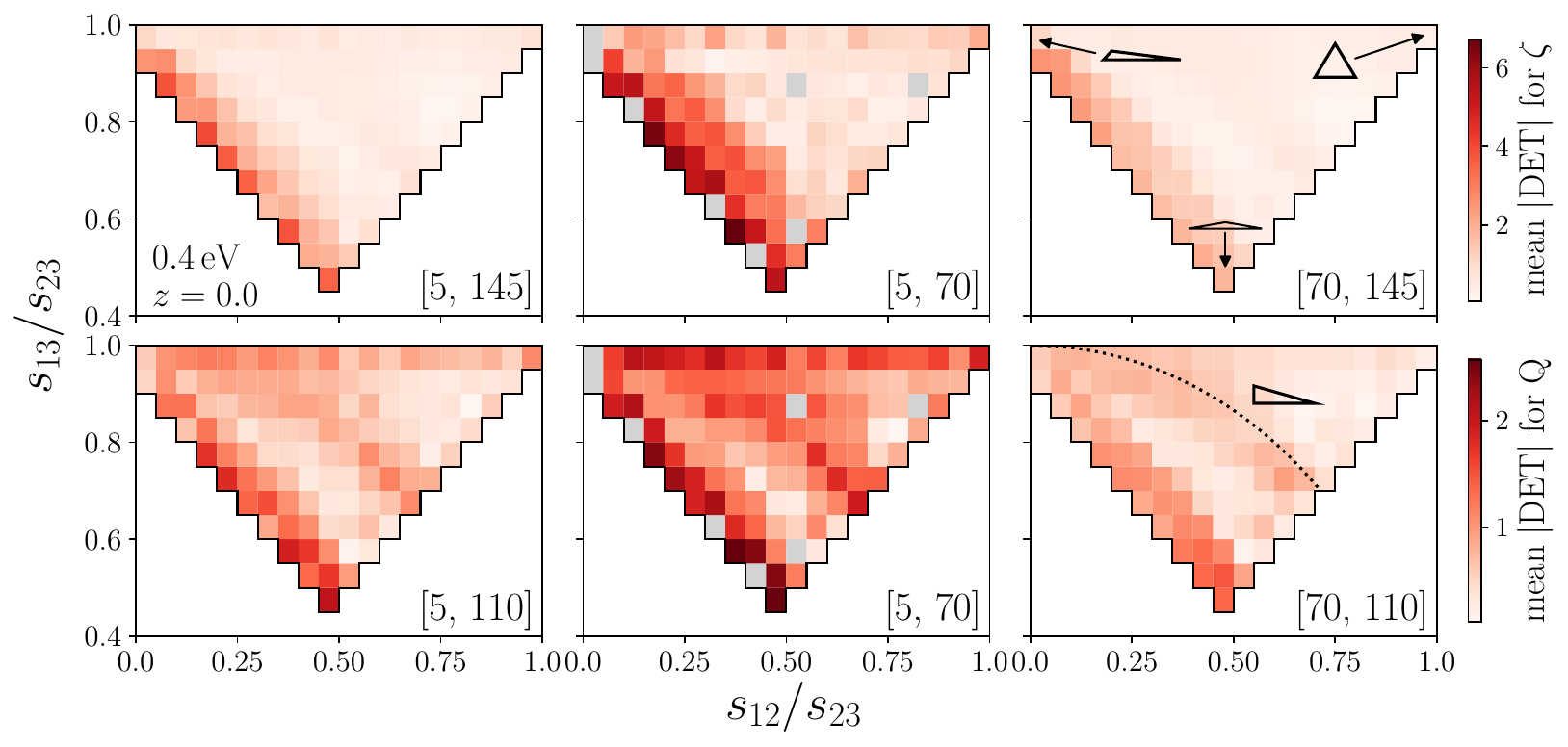}
     \caption{Detectability of the halo connected 3PCF (upper panels) and reduced 3PCF (lower panels) as a function of the triangle shape. The results are reported at $z = 0$ and for $M_\nu = 0.4$ \ev. 
     The triangle shapes are determined by the side ratios $s_{12}/s_{23}$ and $s_{13}/s_{23}$, with $s_{12} \leq s_{13} \leq s_{23}$.
    In each panel, each pixel represents a given triangle shape, where the color shows the absolute value of the detectability averaged over all the triangles available in the all-scales approach with that shape and different sizes. As shown in the legend in the upper right panel, the bottom center, upper right, and upper left parts of the plot contain folded, equilateral, and squeezed triangles, respectively.
    The dotted curve in the lower right panel marks the location of right-angled triangles.
    The white region corresponds to the area in the parameter space where it is not possible to obtain a closed triangle.
    The various columns differ by the range of $s_{23}$ considered, specified in the intervals shown in the bottom right corner (in units of \mpch). 
    Empty bins are colored in gray.}
     \label{fig:shape_plots}
\end{figure*}

Given the overall decrease in neutrino detectability at smaller scales in Fig. \ref{fig:single_scale_chi2_zeta}, it is relevant to quantify the scales at which the statistical significance of the neutrino signal reaches a given threshold, as a function of neutrino mass, redshift, and configuration set, when moving from the largest scales considered in our analysis to progressively smaller ones.
Equations \ref{eq:p-value} and \ref{eq:signif} allow us to convert the $\redchiq(s_{12}, s_{13})$ values into the statistical significance of the signal as a function of the triangle sides $s_{12}$ and $s_{13}$. 
We performed this conversion considering the sets of configurations marked with lines in Fig. \ref{fig:single_scale_chi2_zeta}, which we found to encode most of the information content from neutrinos. 
For this analysis, we identified the characteristic scale $s_{\mathrm{cross}}$ at which, starting from $s_{\max} = 147.5 \, \mpch$ and moving toward smaller scales, the statistical significance first reaches $1\sigma$ and then $3\sigma$ thresholds.
We recall that lower $p$ values correspond to higher significances. For a given combination of $M_\nu$, redshift, and configurations, the scale $s_{\mathrm{cross}}$ can therefore be derived as a function of the statistical significance $N\sigma$ as
\begin{equation}
    s_{\mathrm{cross}}(N\sigma) \equiv \min
    \left\{ s \in [s_{\min}, s_{\max}] \: \Big| \, p(s) \geq p(N\sigma) \right\}\; ,
    \label{eq:scale_crossing}
\end{equation}
\noindent where the scale $s$ can equivalently be considered as either $s_{12}$ or $s_{13}$ (since one side is a function of the other side along the specified lines), $N = 1$ or $3$ for $1\sigma$ or $3\sigma$ significance, respectively, and $p(N\sigma)$ is just the $p$ value corresponding to $N\sigma$ significance. This value therefore represents an upper limit on the scales that need to be included to obtain a given significant detection of neutrinos: the larger this scale, the easier the neutrino effects are detected; the lower it is, the more scales need to be included for a significant detection.

The results are shown in Fig. \ref{fig:summary_zeta} for the connected 3PCF in the form of matrices whose elements report the values of $s_{\mathrm{cross}}$ as a function of $M_\nu$ and $z$.
For isosceles and quasi-isosceles configurations, we set $s = s_{12}$, while for configurations with fixed $s_{12} = 10$ \mpch\, we set $s = s_{13}$.
Overall, a given significance threshold is reached at progressively larger scales for increasing neutrino masses or decreasing redshift, as evident from the increasing values of $s_{\mathrm{cross}}$ moving from the upper matrix row or the leftmost column (associated with $z = 2$ and $M_\nu = 0.1$ \ev, respectively) toward the lower-right matrix element (corresponding to $M_\nu = 0.4$ \ev\ and $z = 0$).
We obtained that only the total masses $M_\nu = 0.2$ \ev\ and $M_\nu = 0.4$ \ev\ are detectable above the $1\sigma$ threshold. 
The mass $M_\nu = 0.2$ \ev\ crosses the $1\sigma$ significance at all redshifts for configurations with $s_{12} = 10$ \mpch, remaining always below $3\sigma$.
The $3\sigma$ threshold is reached for isosceles triangles, but only at the smallest scales probed in our analysis, that is, for $s_{12} \sim 5$ \mpch.
For quasi-isosceles triangles with $\eta = 4$, this mass has a significance above $1\sigma$ for $s_{12} \lesssim 40$ \mpch\ only at $z = 0$.
Our highest mass $M_\nu = 0.4$ \ev\ alone is detectable above $3\sigma$ at all redshifts for isosceles and squeezed configurations.
For these latter, this mass value is already above $1\sigma$ on the largest scale probed in our analysis (147.5 \mpch), unlike for the other masses.
For $\eta = 4$, the signal crosses $3 \sigma$ starting from $z = 1$. 

The analogous results for the reduced 3PCF are shown in Appendix \ref{app:q_scale_dependence}. The overall behavior is that for a fixed neutrino mass, redshift, and configuration, a given statistical significance threshold is reached at smaller scales with respect to $\zeta$, confirming the lower detectability of $Q$ with respect to $\zeta$.

\subsection{Sensitivity of structure shapes to the neutrino signal}
\label{subs:shape_analysis}

We also performed an analysis aimed at quantifying the detectability of massive neutrinos in the 3PCF as a function of triangle shapes.
This is particularly relevant because it leverages information on the morphology of cosmic structures that is not accessible at the two-point level.

The single-scale approach, although effective in examining clustering on a given scale for varying triangle configurations, does not allow an efficient isolation of the contribution from triangles of fixed shape.
For instance, if $s_{12}$ and $s_{13}$ are chosen such that $s_{12} \ll s_{13}$ or $s_{12} \gg s_{13}$, then squeezed triangles are obtained for low and high values of $\theta$. 
The all-scales approach, by ordering the sides of triangles, such that $s_{12} \leq s_{13} \leq s_{23}$ (as detailed in Sect. \ref{subs:dataset_covariance}), allows for a clear shape classification based on the values assumed by two independent side ratios, as we discuss below.

A useful tool for shape analysis is a particular triangular-shaped plot already adopted in Fourier space \citep[e.g.,][]{Takahashi14, Desjacques16_bias_rev, Hahn20, Oddo21}.
In Fig. \ref{fig:shape_plots} we derive its configuration-space counterpart for the first time for the connected and reduced 3PCF. We produced this plot for all combinations of neutrino masses and redshift. In the figure, we show the case $M_\nu = 0.4 \, \ev$ at $z = 0$, which corresponds to the highest detectability.
In this figure, triangle shapes are binned according to the values of the ratios of the smallest and largest side $s_{12}/s_{23}$ and between the intermediate and largest side $s_{13}/s_{23}$. 
The allowed values of $s_{12}/s_{23}$ and $s_{13}/s_{23}$ occupy the triangular-shaped region bounded by the vertices $(s_{12}/s_{23}, s_{13}/s_{23}) =  (0,1), (1/2, 1/2)$ and $(1, 1)$.
Triangles along the leftmost side of this region have elongated shapes ($s_{23} \approx s_{12} + s_{13}$), from squeezed ($s_{12} \ll s_{13} \approx s_{23}$) in the upper left corner to folded ($s_{12} \approx s_{13} \approx 2\,s_{23}$) in the bottom corner, while in the top right corner, we find equilateral triangles. 
We report a visual legend in the upper right panel to facilitate the interpretation of the figure.
It is also interesting to focus on right-angled triangles, residing along the dotted line shown in the bottom right panel (the circular arc with equation $s_{23}^2 = s_{12}^2 + s_{13}^2$).
The color of a given shape bin (i.e., a given pixel) represents the absolute value of the detectability of massive neutrinos defined in Eq. \ref{eq:def_detectability}, averaged over all triangles with that specific shape in a given range of scales.
In the various columns, we show the results for three different ranges of $s_{23}$: the entire range, including all triangles ($5 \, \mpch \leq s_{23} \leq 145 \, \mpch$ for $\zeta$ and $5 \, \mpch \leq s_{23} \leq 110 \, \mpch$ for $Q$), low to intermediate values ($5 \, \mpch \leq s_{23} \leq 70 \, \mpch$ for $\zeta$ and $Q$), and high values ($70 \, \mpch \leq s_{23} \leq 145 \, \mpch$ for $\zeta$ and $70 \, \mpch \leq s_{23} \leq 110 \, \mpch$ for $Q$). 

For $\zeta$, as shown in the upper left plot in the figure, which includes all the triangles considered in our analysis, the most affected regions are those corresponding to the left oblique side of the triangular region, containing elongated triangle shapes transitioning from folded to squeezed. 
The plots in the middle and right columns show that this feature arises from the combined behavior of triangles belonging to the two $s_{23}$ ranges. 
For small to intermediate scales, a comparable contribution to this feature comes from folded and squeezed shapes, also extending to less elongated triangles (i.e., also toward the more central regions of the plot). 
For larger $s_{23}$, the detectability pattern on the left oblique side of the plot concentrates on squeezed configurations.

These changes are naturally explained by considering that higher detectability values in $\zeta$ identify triangles for which at least one side is small.
For triangles with low to intermediate values of $s_{23}$, this is indeed possible for the squeezed and folded shapes, for instance, for triangles like $(s_{12}, s_{13}, s_{23}) \sim  (5,5,10), (10,10,20), (10, 60, 70) \, \mpch$.
Conversely, for $70 \, \mpch \leq s_{23} \leq 145 \, \mpch$, the smallest folded triplet is $(s_{12}, s_{13}, s_{23}) = (35, 35, 70) \, h^{-1} \, \rm{Mpc}$, significantly larger than previous examples. 
Squeezed triangles, instead, can still have low $s_{12}$ values even for the highest $s_{13}$ and $s_{23}$ values, for instance, as in the case of $(s_{12}, s_{13}, s_{23}) \sim (5, 140, 145) \, \mpch$. 

For $Q$, considering all the triangles, we can instead identify at least three regions showing the highest detectabilities: the region associated with squeezed and folded shapes, the isosceles triangles with shape transitioning from squeezed to equilateral (with $s_{13}/s_{23} \approx 1$), and the region of right-angled triangles discussed above.
The fact that this latter region is detectable in $Q$, but not particularly in $\zeta$, can be attributed to the flattening noted on the single-scale $Q$ caused by an increase in $M_\nu$. 
One of the consequences of this flattening indeed is the increase in detectability around $\theta \approx \pi/2$, that is, in correspondence with right-angled triangles.  
The feature present for $s_{13}/s_{23} \approx 1$ is mainly produced by small to intermediate scales, while the neutrino signature on right-angled triangles is present on small to intermediate and large scales.
The reduced 3PCF on elongated triangles is affected by $M_\nu$, with an opposite trend with scale compared with the connected 3PCF: indeed, moving from the smallest to the largest scales, the signal concentrates on folded triangles and decreases for squeezed ones. 

\subsection{Breaking the $M_\nu - \sigma_8$ degeneracy with the 3PCF}
\label{subs:degeneracy}

\begin{figure}
    \centering
    \includegraphics[width=0.92\linewidth]{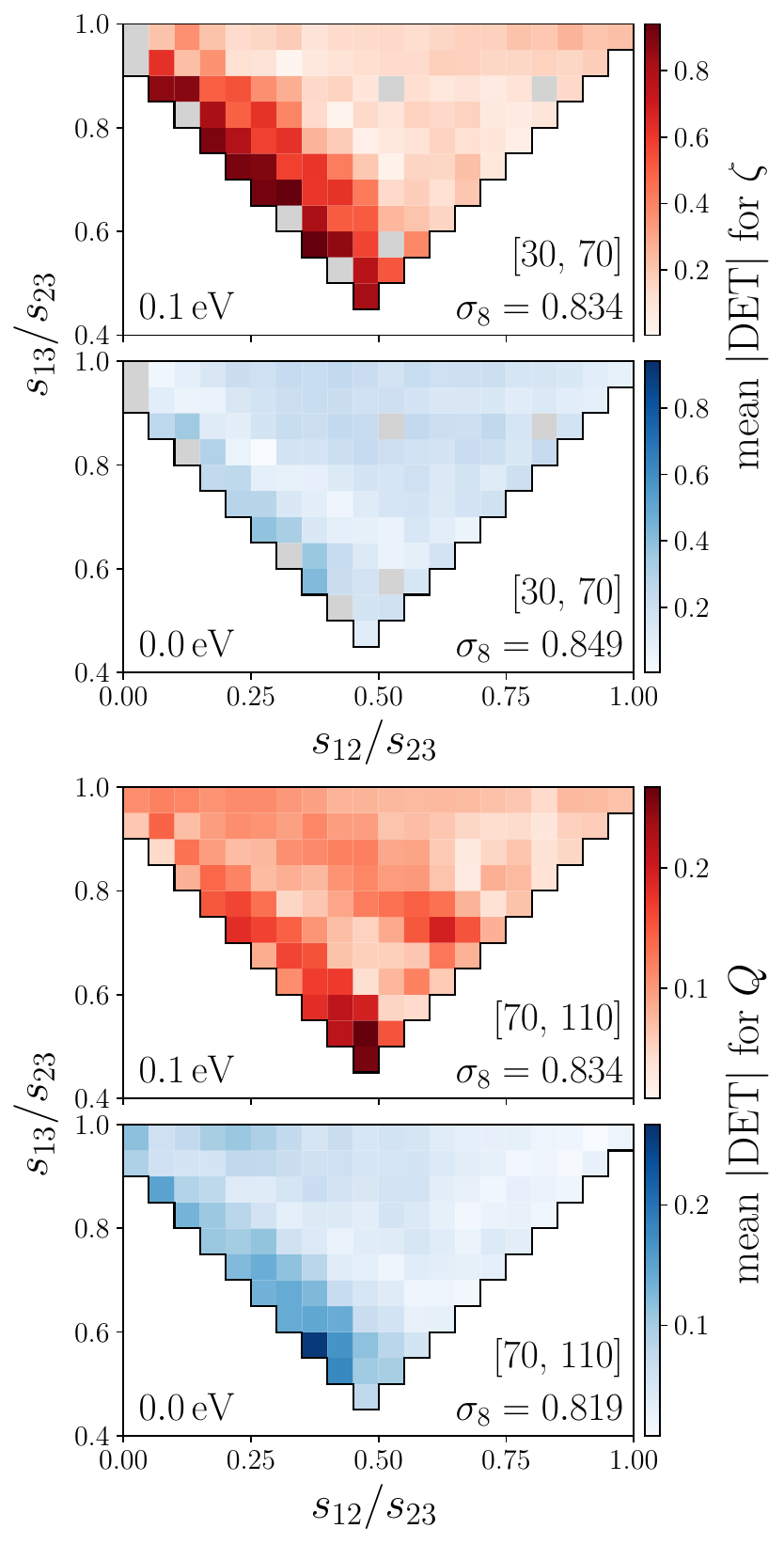}
    \caption{Comparison between the detectability of a variation in $M_\nu$ and $\sigma_8$ from the halo 3PCF as a function of triangle shape.
    We show the results for the simulations at $z = 0$ with $M_\nu = 0.1$ \ev\ and fiducial $\sigma_8 = 0.834$ (red scale color maps), and with $M_\nu = 0$ \ev\ and $\sigma_8 = 0.849$ (upper blue scale color map) and $\sigma_8 = 0.819$ (lower blue scale color map).
    The top and bottom pairs of plots refer to the connected and reduced 3PCF, respectively.
    Triangle shapes are identified by the side ratios $s_{12}/s_{23}$ and $s_{13}/s_{23}$, with $s_{12} \leq s_{13} \leq s_{23}$, as shown in Fig. \ref{fig:shape_plots}.
    In each plot, a given shape bin shows the absolute value of the detectability, averaged over all triangles available in the all-scales approach with that shape and different sizes. 
    We present the results obtained by considering triangles on scales $30 \, \mpch < s_{23} < 70 \, \mpch$ for $\zeta$, and $70 \, \mpch < s_{23} < 110 \, \mpch$ for $Q$.}
    \label{fig:triangle_mnu_s8_deg}
\end{figure}

The suppression of the amplitude of the power spectrum on small scales, caused by neutrino free-streaming, can also be mimicked by a variation in the value of the present-day amplitude of linear matter density fluctuations on a scale of 8 \mpch\ (the parameter $\sigma_8$). 
This causes a well-known degeneracy between the parameters $M_\nu$ and $\sigma_8$ \citep[e.g.,][]{Viel10, VillaescusaNavarro14, Peloso15, VillaescusaNavarro18, Hahn20}.
However, these parameters are not fully degenerate, since the effect of neutrinos is scale dependent and cannot be reduced to a simple $\sigma_8$ renormalization \citep[e.g., see the discussion in][]{Marulli11}. 
Nevertheless, the fact that the imprints of $M_\nu$ and $\sigma_8$ on the power spectrum can even differ by less than 1\% \citep{VillaescusaNavarro18} over a wide range of scales limits the possibility of obtaining precise constraints on $M_\nu$ from two-point statistics alone.

The Fourier-space analysis in \citet{Hahn20} proved that the additional shape information introduced by three-point statistics is promising in breaking this degeneracy, as it shows that the bispectrum exhibits a distinct triangle shape dependence for variations in $M_\nu$ compared to $\sigma_8$.
We employed our framework applied to the massive-neutrino and variable-$\sigma_8$ simulations to perform a shape analysis in configuration space for the 3PCF for the first time. 
In our case, this analysis is meaningful not only for the connected 3PCF, but also for the reduced 3PCF, since the latter is also affected, albeit indirectly, by variations in $\sigma_8$ within our framework. 
For a fixed minimum halo mass, an increase (decrease) in $\sigma_8$ indeed leads to a decrease (an increase) in the halo bias, which in turn has the effect of increasing (decreasing) the amplitude of $Q$ 
\footnote{At linear order and neglecting nonlocal effects, the perturbative relation between the halo density contrast field $\delta_h$ and the underlying matter density contrast $\delta_m$ can be written as $\delta_h = b_1 \delta_m$, where the coefficient $b_1$ is called linear bias \citep{Kaiser84, Bardeen86, ColeKaiser89, MoWhite96, ShethTormen99, Desjacques16_bias_rev}.
The amplitude of the reduced 3PCF of the halos scales with the linear bias $b_1$ as $1/b_1$ \citep[see Eq.13 in][]{Gaztanaga09}.}.

We compared the triangular plots (for a detailed description of the structure of these plots, we refer to Sect. \ref{subs:shape_analysis}) for the simulations with massive neutrinos and variable-$\sigma_8$. 
In Fig. \ref{fig:triangle_mnu_s8_deg} we show two particularly informative cases: the detectability for simulations $(M_\nu, \sigma_8) = (0.1\,\ev, 0.834)$ compared to $(M_\nu, \sigma_8) = (0.0\,\ev,\,0.849)$ in the case of $\zeta$ for triangles with $30 \, \mpch \leq s_{23} \leq 70 \, \mpch$, and with simulations $(M_\nu, \sigma_8) = (0.0\,\ev, 0.819)$ in the case of $Q$ for triangles with $70 \, \mpch \leq s_{23} \leq 110 \, \mpch$. 
In Appendix \ref{app:quant_mnu_sigma8} we extend the discussion to the remaining neutrino masses and scales considered in this analysis. 
We considered the mass $M_\nu = 0.1\, \ev$ to facilitate visual inspection, as its detectability is comparable to that of the variable-$\sigma_8$ simulations, and we selected the values $\sigma_8 = 0.849$ and $0.819$ for $\zeta$ and $Q$, respectively, because they produce a steepening of $\zeta$ (its amplitude increases) and a flattening of $Q$ (due to the previously discussed effect of the halo bias), potentially mimicking the imprint of neutrinos discussed in Sect. \ref{subs:singlescale_analysis}. 
We note that this represents a lower limit on the possibility of discriminating between the two signals; clearly, the results obtained for higher neutrino masses will be significantly more distinguishable.

In $\zeta$, we observe that while the detectability values in the massive-neutrino and variable-$\sigma_8$ cases are similar in the upper right region of the plots, they differ significantly for elongated triangles (left edge), with the shape-dependent effect induced by massive neutrinos being much more pronounced.
In $Q$, the most evident difference is that the feature produced by massive neutrinos at the location of right-angled triangles is absent for the considered variation in $\sigma_8$. 
Furthermore, the shape dependence for folded or near-folded triangles (lower corner) appears different in the two cases.
A more quantitative analysis presented in Appendix \ref{app:quant_mnu_sigma8} shows that at $z=0$ for $\zeta$, the average detectability on elongated triangles reaches a $\gtrsim 7 \sigma$ discrepancy between massive-neutrino and varying-$\sigma_8$ simulations, remaining $\lesssim 2 \sigma$ for the other shapes; for $Q$, the discrepancy is $\gtrsim 3 \sigma$ for right-angled and elongated triangles, lying within $\sim 1 \sigma$ in the other cases. 

Hence, we demonstrate that the three-dimensional shape information of cosmic structures encoded in the 3PCF (unlike the 2PCF, which only captures scale information) is effective in disentangling the effects of massive neutrinos from $\sigma_8$ rescalings. 
Our analysis indeed shows that the shape-dependent imprint of massive neutrinos on the 3PCF cannot be replicated by just rescaling $\sigma_8$, and it identifies the configurations that cause this behavior.

\section{Conclusions}
\label{sec:conclusions}

We studied the effect of massive neutrinos on the halo three-point statistics in configuration space for the first time by using $2\,000$ $N$-body simulations from the \textsc{Quijote} suite \citep{quijote_paper}.
We considered simulations at redshifts $z = 0,1$, and 2, characterized by values of the sum of neutrino masses $M_\nu = 0.0, 0.1, 0.2$, and $0.4$ \ev.

We estimated the isotropic connected 3PCF $\zeta$ and the reduced 3PCF $Q$ (Eq. \ref{eq:q_definition}), and  we made all the measurements publicly available.
To estimate the latter, we also measured the halo 2PCF of our simulations. 
We ran our measurements with the code \texttt{MeasCorr} \citep{Farina26}, probing a wide range of scales, from 1 to $150 \, h^{-1} \, \mathrm{Mpc}$ for the 2PCF and from 5 to $145 \, h^{-1} \, \mathrm{Mpc}$ for the 3PCF, thanks to the implementation of the \citet{SlepianEisenstein15} SHD estimator.
We numerically estimated the covariance matrices of the considered statistics from an additional set of $2\,000$ $\Lambda$CDM simulations from the \textsc{Quijote} suite itself.

We developed a framework to quantify the neutrino signal in comparison with the fiducial massless case by introducing detectability metrics depending on single and multiple configurations, and on their correlation. 
We applied this framework to determine the neutrino detectability in a volume of 10 \gpchc\ representative of an ideal redshift bin of a stage-IV survey by rescaling the estimated covariance.

We also used our framework to study whether $\zeta$ and $Q$ can provide distinct features that allow us to break the degeneracy between $M_\nu$ and $\sigma_8$ that affects two-point statistics. 
To do this, we complemented our previous dataset (in which $\sigma_8$ was fixed at the fiducial value 0.834) with measurements performed on three sets of 500 \textsc{Quijote} massless neutrino simulations each, with $\sigma_8 = 0.819, 0.834$, and 0.849. The main results of our analysis are listed below.   

\begin{itemize}
    \item 
    For $\zeta$, we found that the effect of neutrinos is stronger for isosceles triangles below $\sim 30 \, \mpch$ and for quasi-isosceles and squeezed triangles. 
    For $Q$, we observe a maximum for isosceles triangles, with detectabilities overall lower than for $\zeta$. 
    In correspondence with these configurations, for increasing $M_\nu$, we found that the concavity of $\zeta$ as a function of the angle between two fixed sides grows, while $Q$ flattens.

    \item 
    The signal from massive neutrinos increases with decreasing scale and redshift, following the evolution of nonlinearities.
    In particular, it is interesting to note that neutrino masses $M_\nu = 0.4 \, \ev$ can be detected at a $3\sigma$ significance already at intermediate or large triangle scales (above $\sim 50 \, \mpch$, with values depending specifically on the configuration), while for $M_\nu = 0.2 \, \ev$, the $3\sigma$ significance is only reached by pushing the configurations to small nonlinear scales ($\sim 5-10 \, \mpch$), which are currently quite difficult to model.
     
    \item 
    We found that massive neutrinos predominantly affect the filaments of the cosmic web.
    The enhanced sensitivity of filamentary configurations to massive neutrinos was recently reported also by \citet{Pal25}, through an analysis of the redshift-space bispectrum multipoles.
    Although most of the signal is on small scales due to free-streaming, we showed that BAO scales are also affected by their presence, but these effects are largely undetectable within the volume probed by stage-IV surveys.

    \item 
     By studying the effect of neutrino masses on triangle shapes, we found in $\zeta$ the strongest signal for elongated triangles, confirming that filamentary shapes contain the strongest signal, with a preference for squeezed triangles when larger scales are included. 
    For $Q$, we found that an additional source of signal with respect to $\zeta$ is represented by right-angled triangles.

    \item 
    We found that the shape dependence of $\zeta$ and $Q$ is affected differently by variations in $M_\nu$ and $\sigma_8$. 
    In particular, the main differences were found for elongated triangles in $\zeta$ and for elongated and right-angled ones for $Q$. 
    The average detectability shows discrepancies of $\gtrsim 7\sigma$ on elongated triangles between massive-neutrino and varying-$\sigma_8$ simulations, while remaining below $\lesssim 2\sigma$ for other configurations. For $Q$, right-angled and elongated triangles yield discrepancies of $\gtrsim 3\sigma$, whereas the remaining shapes are consistent within $\sim 1\sigma$.
\end{itemize}

This work showed for the first time in the literature that the 3PCF might be employed as a cosmological tool beyond the standard $\Lambda$CDM model, in particular, for probing neutrino masses. 
The extent to which this potential can be effectively unlocked will significantly depend on the smallest scales that can be accurately modeled, a challenging task due to the increasing effect of nonlinearities.
For reference, scales below $\sim 20$ \mpch\ are currently excluded from the validity range of perturbative models \citep[e.g.,][]{SlepianEisenstein15_model, SlepianEisenstein17_model,
KamalinejadSlepian25, Farina26}.
Moreover, the maximum amount of information from clustering studies can be extracted by jointly analyzing lower- and higher-order statistics, as well as by including different observational probes, for example, CMB. We focused extensively on the 3PCF, with the aim of exploring these approaches in future works. 

Our analysis also demonstrated that the reduced 3PCF enables the identification of neutrino-induced signatures in structures that are weakly sensitive to their effect in the connected 3PCF (e.g., right-angled triangles).
Moreover, the reduced 3PCF has the advantage that its modeling is independent of $\sigma_8$.
Nevertheless, due to the overall lower detectability of the reduced 3PCF compared to the connected 3PCF, a more synergistic exploitation of the former together with the latter will require cosmological volumes larger than those targeted by current surveys. 

This paper is the first of an exploratory research program aimed at quantifying the neutrino information content encoded in the 3PCF. 
In future analyses, we plan to investigate the full-shape dependence of the 3PCF on cosmological parameters (including the sum of neutrino masses) to quantify its constraining power and combine the information provided by higher-order statistics with two-point statistics in a joint likelihood analysis, to maximize the scientific return from galaxy clustering in view of forthcoming data.
In this perspective, this work is directly relevant in preparation of the data releases from stage-IV surveys such as Euclid \citep{Euclid_definition_laureijs, Euclid_definition}, DESI \citep{DESI_definition}, 4MOST \citep{4MOST_def}, and the Nancy Grace Roman Space Telescope \citep{nancy_grace}, and its potential will be further enhanced by the even larger volumes covered by future stage-V facilities, such as the proposed Wide-field Spectroscopic Telescope \citep[WST;][]{wst_white_paper}, which aims to map the galaxy distribution up to $z \sim 5.5$, providing unprecedented statistics.

\begin{acknowledgements}
We thank the referee for the constructive comments, which helped to improve the clarity of our paper.
We thank Francisco Antonio Villaescusa-Navarro for supporting the public release of our measurements as part of the \textsc{Quijote} suite data products, and Antonio Farina and Marco Baldi for useful discussions. 
AL acknowledges the use of computational resources provided by the ``Open Physics Hub'' cluster at the Department of Physics and Astronomy of the University of Bologna.
MM and MG acknowledge the financial contribution from the grant PRIN-MUR 2022 2022NY2ZRS 001 “Optimizing the extraction of cosmological information from Large Scale Structure analysis in view of the next large spectroscopic surveys” supported by Next Generation EU.
MM acknowledges support from the grant ASI n. 2024-10-HH.0 “Attività scientifiche per la missione Euclid – fase E”.
\end{acknowledgements}

\bibliographystyle{aa} 
\bibliography{biblio}

\begin{appendix}
\section{Neutrino detectability in the 3PCF multipoles}
\label{app:multipoles}

\begin{figure*}
\centering
   \includegraphics[width= 17 cm]{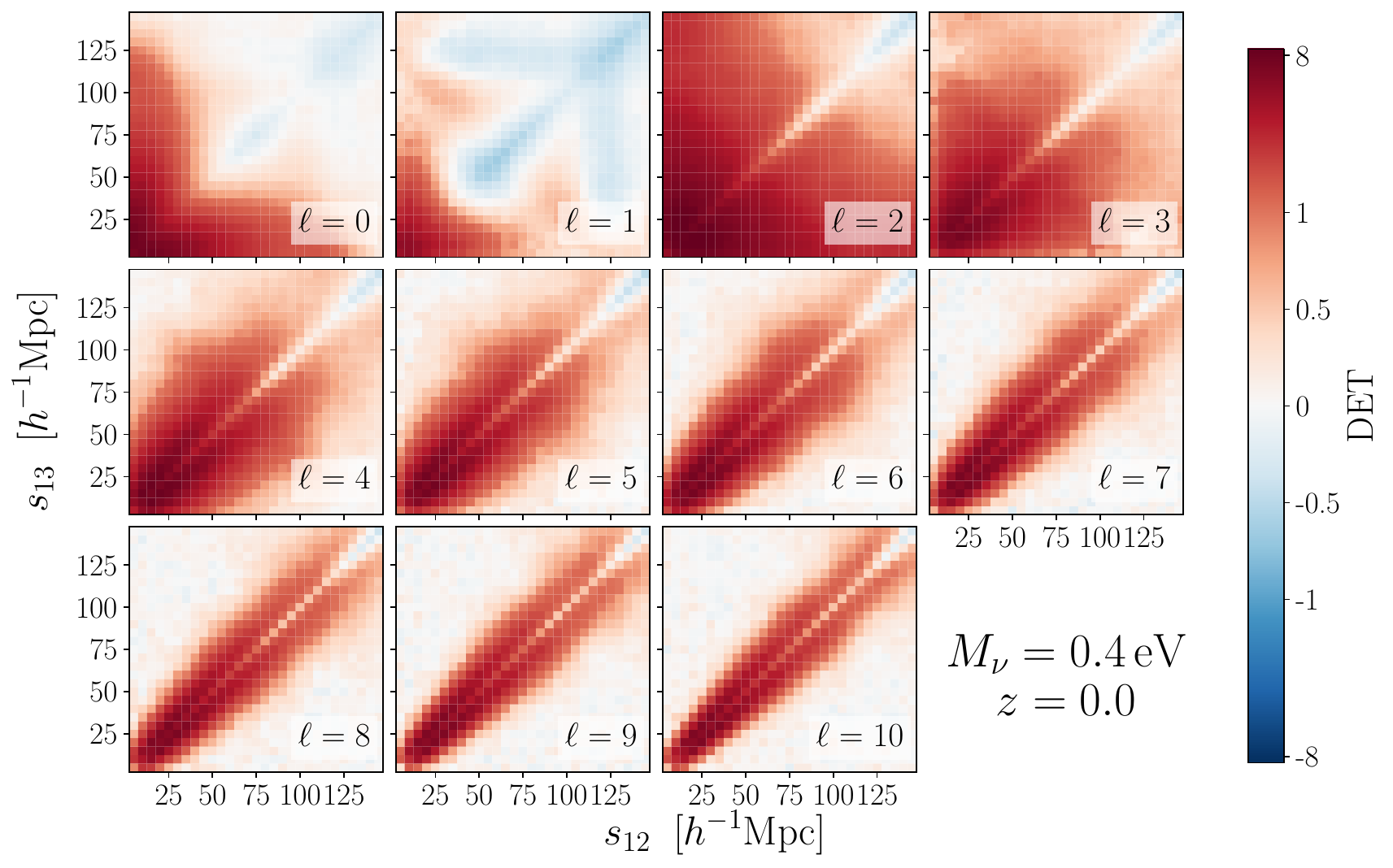}
     \caption{Detectabilities of neutrino masses (computed with Eq. \ref{eq:def_detectability}) for the multipoles of the connected 3PCF $\zeta_\ell(s_{12}, s_{13})$. This figure shows the results for the case of $M_\nu = 0.4 \, \rm{eV}$ at $z=0$.}
     \label{fig:multipoles_detectability}
\end{figure*}

We studied how massive neutrinos affect the multipoles of the connected 3PCF by comparing the detectability of different neutrino masses with the massless case.
To avoid repetition, we do not show the analysis for all the possible combinations of redshifts and neutrino masses, but we plot in Fig. \ref{fig:multipoles_detectability} the results obtained for $M_\nu = 0.4$ \ev\ at $z = 0$ as a representative case.
We show the detectabilities through a 2D colormap for each multipole $\ell$ (in the measured range $0 \leq \ell \leq 10$), where a given pixel corresponds to a binned triangle side pair $(s_{12}, s_{13})$.

The fact that neutrinos mostly affect small, nonlinear scales is evident by inspecting the figure, where the highest, positive values of detectability are mostly concentrated on small scales for all the measured $\ell$. 
For $\ell = 0$, whose corresponding monopole $\zeta_0$ only contributes to the offset of $\zeta$, squeezed triangles (corresponding to the regions adjacent to the bottom and left sides of each colormap, far from the bottom-left corner) also show a positive detectability, decreasing for increasing scale. 

The monopole $\ell = 0$ and the dipole $\ell =1$ change their sign, exhibiting a mildly negative detectability (for which, in both cases $-1 < \detec < 0$), on a large region of the explored $(s_{12}, s_{13})$ area.
For $\ell = 0$, this feature is mostly located along the diagonal region (i.e., for isosceles triangles), divided into two features with $50 \, \mpch \lesssim s_{12}  \lesssim 95 \,  \mpch$ and $s_{12} \gtrsim 105 \, \mpch$, separated across BAO scales.
For $\ell = 1$, it shows a fork-shaped pattern, made of three features, one located on the diagonal on scales $s_{12} \gtrsim 40 \, \mpch$, and the other ones on scales $s_{12} \gtrsim 25 \, \mpch$ when $115 \, \mpch \lesssim s_{13} \lesssim 135 \, \mpch$ and analogously with $s_{12}$ and $s_{13}$ interchanged. 
In the latter two cases, the change in sign observed at fixed $s_{13} \sim 105 \, \mpch$ (or equivalently at fixed $s_{12} \sim 105 \, \mpch$) is driven by the effect of the BAO on the dipole, as shown by the model of \citet{SlepianEisenstein17_model}. This multipole is sourced by the gradient of the density field; therefore, it exhibits a zero crossing at BAO scales, since the acoustic feature corresponds to local maxima in the density distribution.

The multipole that, in the presence of massive neutrinos, shows the most significant departures from the fiducial cosmology is the quadrupole $\zeta_2$, which imprints the characteristic parabolic shape to $\zeta$ as a function of the angle between two fixed sides, as evident in Fig. \ref{fig:single_scales_zeta}. 
For $\zeta_2$, on scales $s_{12}, s_{13} \lesssim 25 \, \mpch$ the detectability for $M_\nu = 0.4 \, \ev$ is around 8 and for squeezed configurations it remains $\gtrsim 2$ even for the largest scales probed in our measurement, vanishing only for triangles with large $s_{12}$ and $s_{13}$.
Starting from $\ell = 3$, the highest-detectability pattern progressively concentrates on quasi-isosceles configurations, decreasing with increasing $s_{12}$ and $s_{13}$.

Interestingly, for isosceles configurations and $\ell \geq 2$, the detectability patterns show a local minimum.
We find that this drop is driven by a local maximum in the standard deviations of all multipoles.
Although this maximum is also present for $\ell = 0$ and $\ell = 1$, the complex detectability pattern of these two multipoles prevents it from being evident in the first two panels of Fig. \ref{fig:multipoles_detectability}.
Our analysis suggests that this feature originates from autocorrelation within the same radial bin, which occurs for the 3PCF multipoles of isosceles triangles estimated with the \citet{SlepianEisenstein15} algorithm, as in our case. 
In the presence of autocorrelation, the expected variance is larger than in the case of cross-correlation between two different radial bins. 

\section{Single-scale analysis for the reduced 3PCF}
\label{app:q_single-scale}

\begin{figure*}
\centering
   \includegraphics[width= 17 cm]{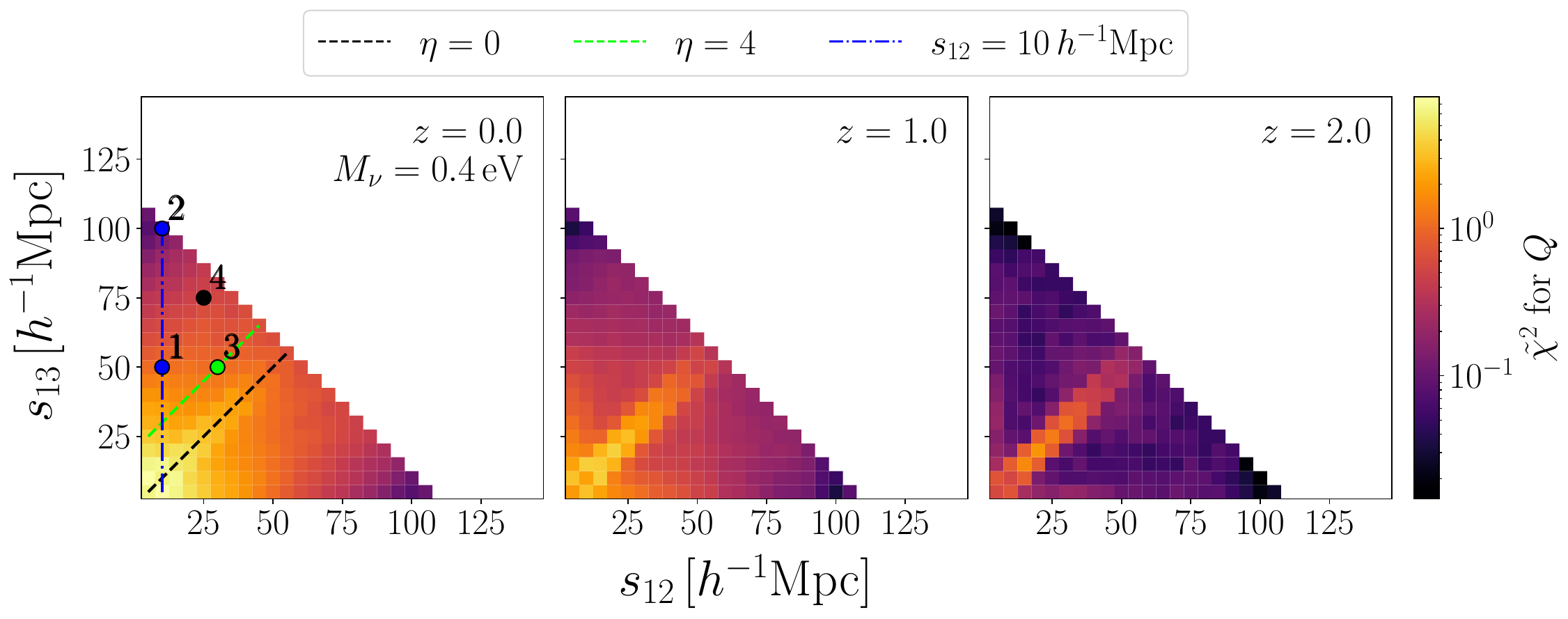}
     \caption{Same as Fig. \ref{fig:single_scale_chi2_zeta} but for the reduced 3PCF $Q$.  
     The white region in each panel corresponds to the configurations excluded by the constraint $s_{12} + s_{13} \leq 110$ \mpch.
     The lines are the same as those overplotted in Fig. \ref{fig:single_scale_chi2_zeta}, for easier comparison with $\zeta$.
     The $(s_{12}, s_{13})$ configurations marked with a numbered circle correspond, in increasing order from 1 to 4, to $(s_{12}, s_{13}) = (10,50), (10,100), (30,50)$, and $(25,75)$ \mpch. 
     The single-scale $Q$ for these scales is plotted in Fig. \ref{fig:single_scale_q}. 
     }     \label{fig:single_scale_chi2_q}
\end{figure*}

In Fig. \ref{fig:single_scale_chi2_q} we show the values of the parameter $\redchiq$ (Eq. \ref{eq:single-scale_chi_det}) for the reduced 3PCF, computed for the largest neutrino mass $M_\nu = 0.4 \, \ev$ at redshifts $z = 0,1$, and 2.
We exclude all the configurations $(s_{12}, s_{13})$ for which the third side $s_{23}$ can cross the value $110 \, \mpch$, to avoid singularities of $Q$.

The imprint of neutrinos is mainly concentrated in the $(s_{12}, s_{13})$ bins located along the diagonal, and in the two adjacent ones. 
The values of $\redchiq$ decrease with increasing distance from the diagonal region. 
Similarly to $\zeta$, the signal increases going to small scales and for decreasing redshift. 

For consistency with Fig. \ref{fig:single_scales_zeta}, we show on the figure the same lines marking the locations of the isosceles, squeezed, and quasi-isosceles configurations with $\eta = 4$.
\begin{figure*}
\centering
   \includegraphics[width= 17 cm]{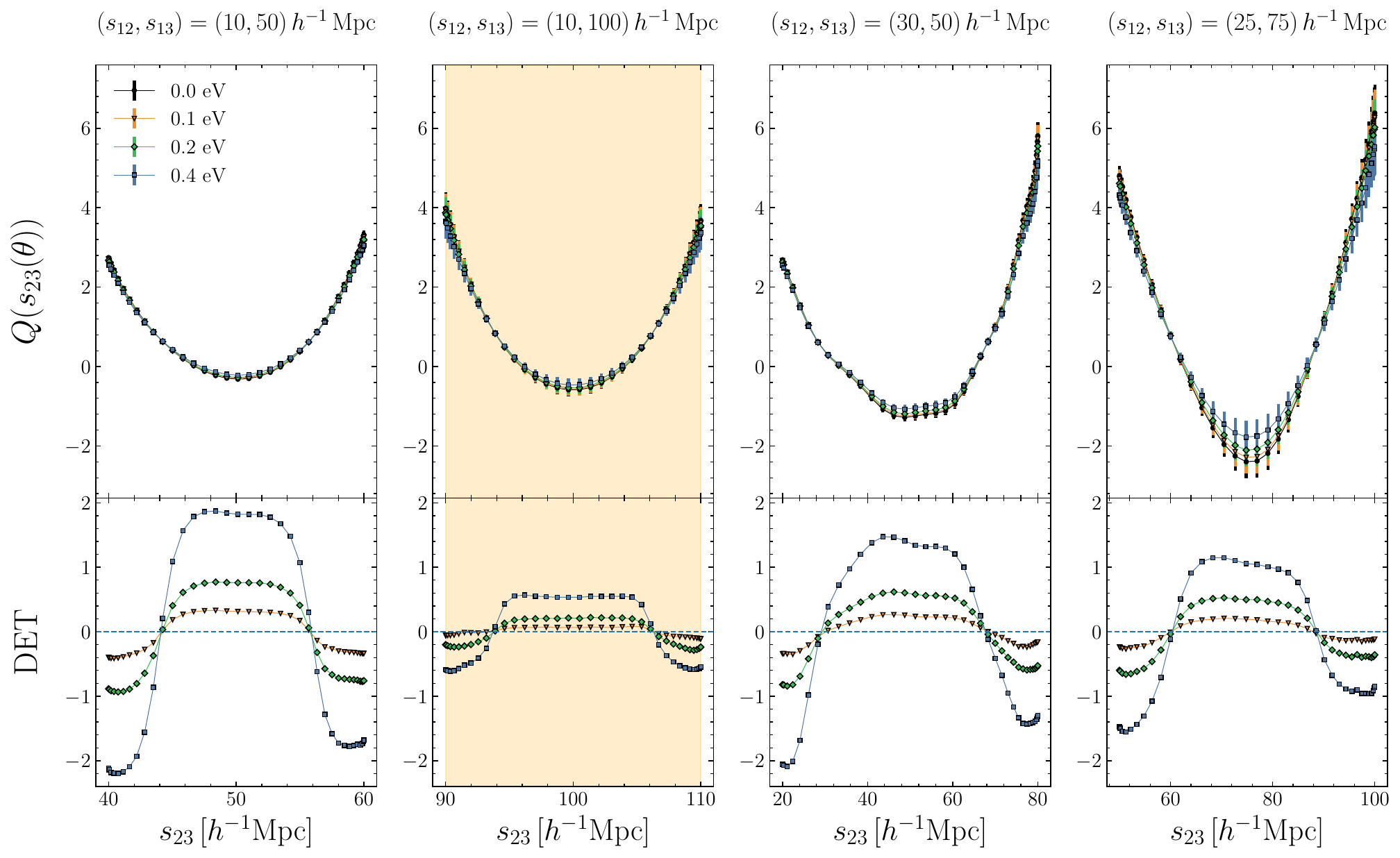}
     \caption{Single-scale reduced 3PCF $Q$ (upper plot of each panel) for the $(s_{12}, s_{13})$ configurations selected in Fig. \ref{fig:single_scale_chi2_q} and indicated at the top, as a function of the third side $s_{23}$. We show here the results for $z = 0$, with a different color for each of the considered neutrino masses, as shown in the legend.
     The lower plots show the corresponding detectabilities as a function of $s_{23}$ (Eq. \ref{eq:def_detectability}), with the same color coding as $Q$.
     A blue dashed line marks the zero detectability level.
     The orange shaded area shows the region $90 \, \mpch \leq s_{23} \leq 110 \, \mpch$, corresponding to the expected location of the BAO peak.
     }
     \label{fig:single_scale_q}
\end{figure*}
Consistent with the applied scale cut, we select specific configurations, corresponding to $(s_{12}, s_{13}) = (10, 50), (10, 100), (30, 50)$, and $(25,75) \, \mpch$  along and off these lines (for comparison) to inspect the angular behavior of $Q$ visually.
We show the corresponding reduced 3PCFs as a function of the third side $s_{23}$ for all the values of $M_\nu$ at $z = 0$, and their associated detectability with respect to $M_\nu = 0 \, \ev$, in Fig. \ref{fig:single_scale_q}. 

As anticipated in Sect. \ref{subs:definitions}, the order of magnitude of $Q$ does not change when varying the configurations $(s_{12}, s_{13})$ and remains approximately of order unity, unlike $\zeta$ in Fig. \ref{fig:single_scales_zeta}, which instead varies by about one order of magnitude across the selected scales and decreases with increasing scale.
The concavity of $Q$ decreases as $M_\nu$ increases, causing a slight flattening of the function with respect to the measurements for the fiducial massless cosmology.
This flattening is reflected in the detectability profile, producing $s_{23}$-dependent trends with a facing-down concavity: the detectability becomes negative for filamentary structures, i.e., at the edges of the $s_{23}$ range, and positive for central values.
These configurations are the most affected by neutrinos, as they exhibit the largest absolute detectability values.
In particular, note that the detectability at intermediate configurations remains relatively stable, with a magnitude comparable to that at the extremes.
The flattening of $Q$ with increasing $M_\nu$ also causes each function measured in the massive-neutrino simulations to intersect the corresponding massless curve at two points, driving the detectability to zero.
By inspecting the detectabilities in Fig. \ref{fig:single_scale_q}, this occurs, regardless of the value of $M_\nu$, just below $\sim 1/4$ and just above $\sim 3/4$ of the $s_{23}$ range.

Due to our scale cut, we do not identify any BAO feature in $Q$. 
This occurs because the $s_{23}$ range is entirely dominated by BAO scales.
This is clearly shown in Fig. \ref{fig:single_scale_q}, where the orange shaded area, corresponding to the expected location of possible BAO features, spans the entire $s_{23}$ range for $(s_{12}, s_{13}) = (10, 100) \, \mpch$.  
This implies that any potential BAO signal is spread over the entire range of $s_{23}$ values, and therefore does not appear as a visible, localized peak.

\section{Scale dependence of the neutrino signal for the reduced 3PCF}
\label{app:q_scale_dependence}

\begin{figure*}
\centering
   \includegraphics[width= 17 cm]{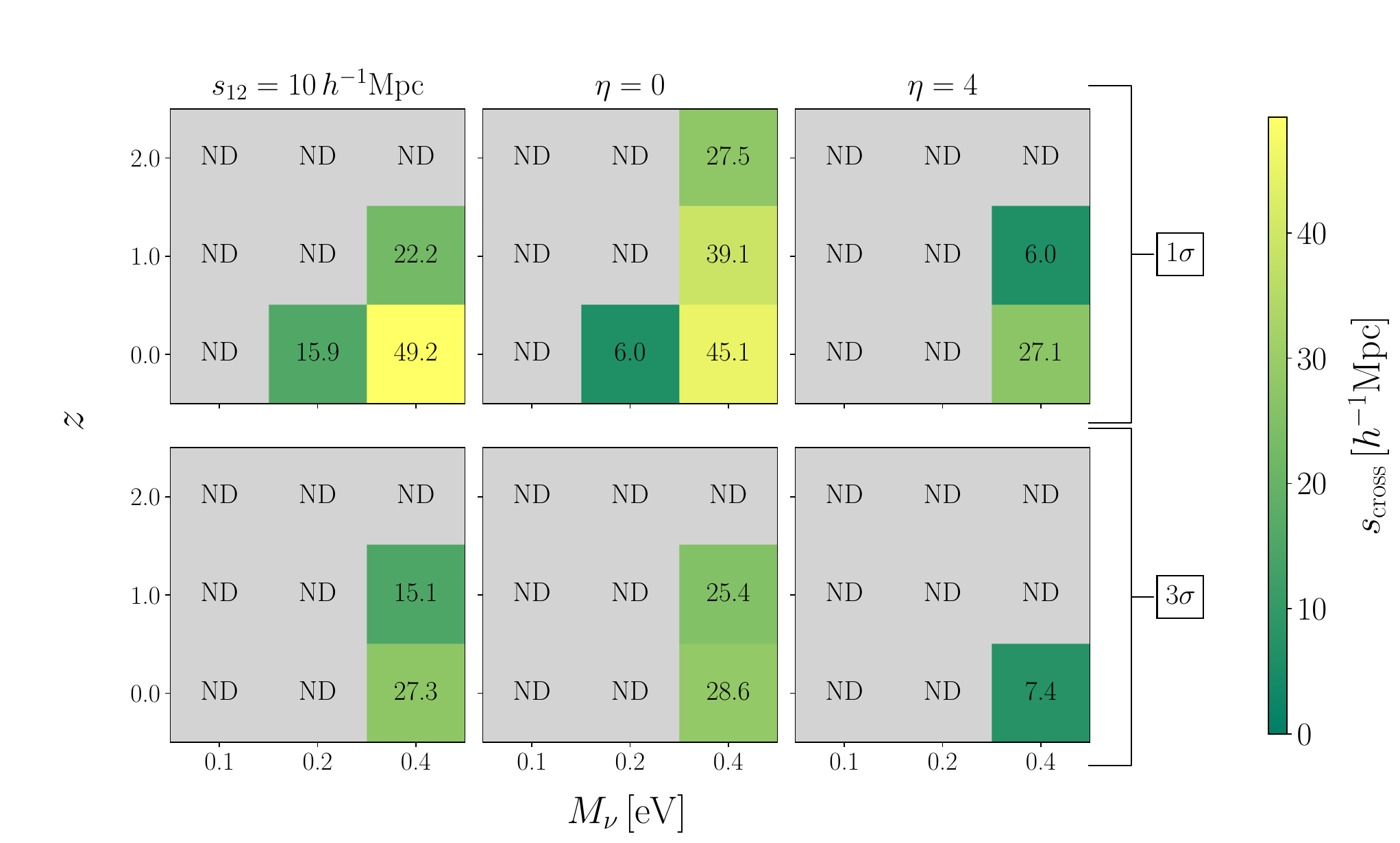}
     \caption{Same as Fig. \ref{fig:summary_zeta} but for the reduced 3PCF $Q$.}
     \label{fig:summary_q}
\end{figure*}

We assessed for $Q$, as we did for $\zeta$ in Sect. \ref{subs:scale_analysis}, the significance of the massive-neutrino signal relative to the fiducial cosmology as a function of scale, for specific configurations.
To do so, we applied the same procedure detailed in Sect. \ref{subs:scale_analysis} to the $\redchiq$ values of the configurations $(s_{12}, s_{13})$ along the lines shown in Fig. \ref{fig:single_scale_chi2_q}. 

Figure \ref{fig:summary_q} shows the scales $s_{\mathrm{cross}}$ (already defined in Eq. \ref{eq:scale_crossing}) on which the statistical significance of the signal produced by a given neutrino mass $M_\nu$ at a given redshift becomes higher, when moving from large to small scales, than $1\sigma$ and $3\sigma$ significance.
We follow the same convention of Fig. \ref{fig:summary_zeta}, by setting in Eq. \ref{eq:scale_crossing} $s = s_{12}$ for isosceles and quasi-isosceles configurations with $\eta = 4$ (being $s_{13}$ given by, respectively, $s_{13} = s_{12}$ and $s_{13} = s_{12} + 4 \times \Delta s$ \mpch), and $s = s_{13}$ for triangles with $s_{12} = 10 \, \mpch$. 

Since, as visible in Fig. \ref{fig:single_scale_chi2_q}, the signal increases moving toward smaller scales while remaining globally weaker than in the case of $\zeta$, the scales at which it reaches a given significance are smaller for $Q$ than for $\zeta$. 
This is evident by directly comparing the values of $s_{\mathrm{cross}}$ for each matrix element in Fig. \ref{fig:summary_q} with respect to Fig. \ref{fig:summary_zeta}.
This implies that cosmological volumes larger than $10 \, \gpchc$ are required for $Q$ to achieve detectability levels comparable to those shown for $\zeta$.

The figure also shows that restricting the analysis to scales below $110 \, \mpch$, i.e., those considered for $Q$ in our study, excludes only a negligible fraction of the neutrino signal in a volume of $10 \, \gpchc$, given that Fig. \ref{fig:summary_q} indicates that the bulk of the signal is located below $50 \, \mpch$ (there are no printed values of $s_{\mathrm{cross}}$ exceeding this threshold).
Specifically, only the masses $M_\nu = 0.2 \, \ev$ and $0.4 \, \ev$ are detectable above $1\sigma$.
In particular, cosmic neutrinos with a total mass of $M_\nu = 0.2 \, \ev$ are detectable above $1\sigma$ only at $z= 0$ (but never above $3\sigma$) for isosceles configurations and for $s_{12} = 10 \, \mpch$, up to scales of $\sim 6 \, \mpch$ and $16 \, \mpch$, respectively.
The mass $M_\nu = 0.4 \, \ev$ is detectable above $1\sigma$ at all redshifts only for $\eta = 0$, and at $z = 0$ and $z = 1$ for squeezed and quasi-isosceles configurations.
For these redshift values, squeezed and isosceles triangles also exceed the $3\sigma$ detectability on scales between $\sim 15$ and $30 \, \mpch$, while only at $z = 0$ for $s_{\mathrm{cross}} \sim 7 \, \mpch$ in the case of quasi-isosceles triangles.
At $z=2$, the statistical significance never reaches $3\sigma$, regardless of neutrino mass, redshift, and configuration.

\section{Quantifying the information content of different triangle shapes with respect to variations in $M_\nu$ and $\sigma_8$}
\label{app:quant_mnu_sigma8}

Here, we complement the analysis presented in Sect. \ref{subs:degeneracy} by quantifying the capability of different triangle shapes to disentangle the effect of variations in the neutrino mass, $M_\nu$, and in $\sigma_8$ on the connected and reduced 3PCF.
The triangular plots shown in Fig. \ref{fig:triangle_mnu_s8_deg} indicate that the largest differences are concentrated in regions of the $(s_{12}/s_{23}, s_{13}/s_{23})$ plane that lie approximately parallel to the left side of the triangular domain (elongated or right-angled triangles). 
For this reason, we investigated the detectability in the two different sets of simulations by estimating the average $|\detec|$ along directions parallel to the left oblique side, which therefore is expected to provide a more informative way to probe shape-dependent effects.

\begin{figure}
    \centering
    \resizebox{\hsize}{!}{\includegraphics{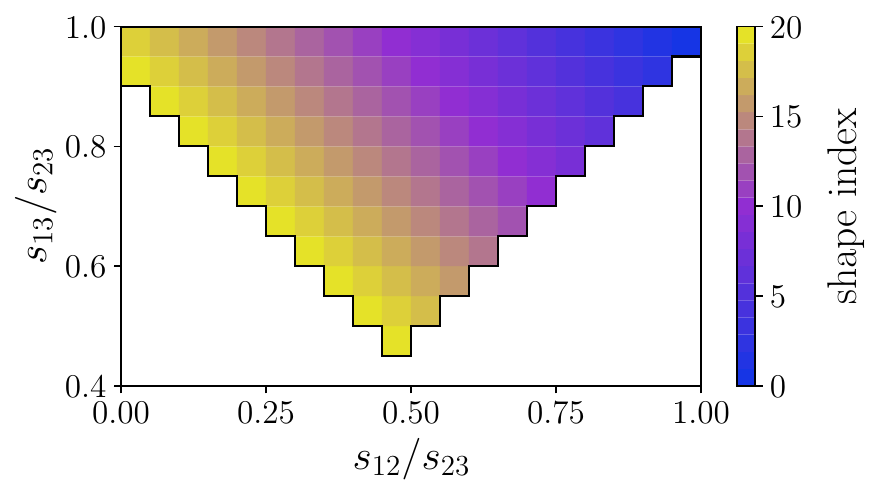}}
    \caption{Legend illustrating the definition of the ``shape index'' adopted in our analysis for Fig.~\ref{fig:diff_shape_det}. 
    A given shape index groups all shape bins lying along a line parallel to the left oblique side, increasing from the top-right corner (equilateral triangles) towards the left side (elongated triangles), as indicated in the colorbar.}
    \label{fig:legend_slicing}
\end{figure}

To this end, we divide our diagram into bins of different shapes (from equilateral to squeezed), which we call ``shape index'', as shown in Fig. \ref{fig:legend_slicing}. 
We assign a shape index value of 0 to the top-right bin, which contains equilateral triangles, and increase the shape index towards the left side, which corresponds to a value of 20 for the binning adopted in this work.

\begin{figure*}
\centering
   \includegraphics[width= 17 cm]{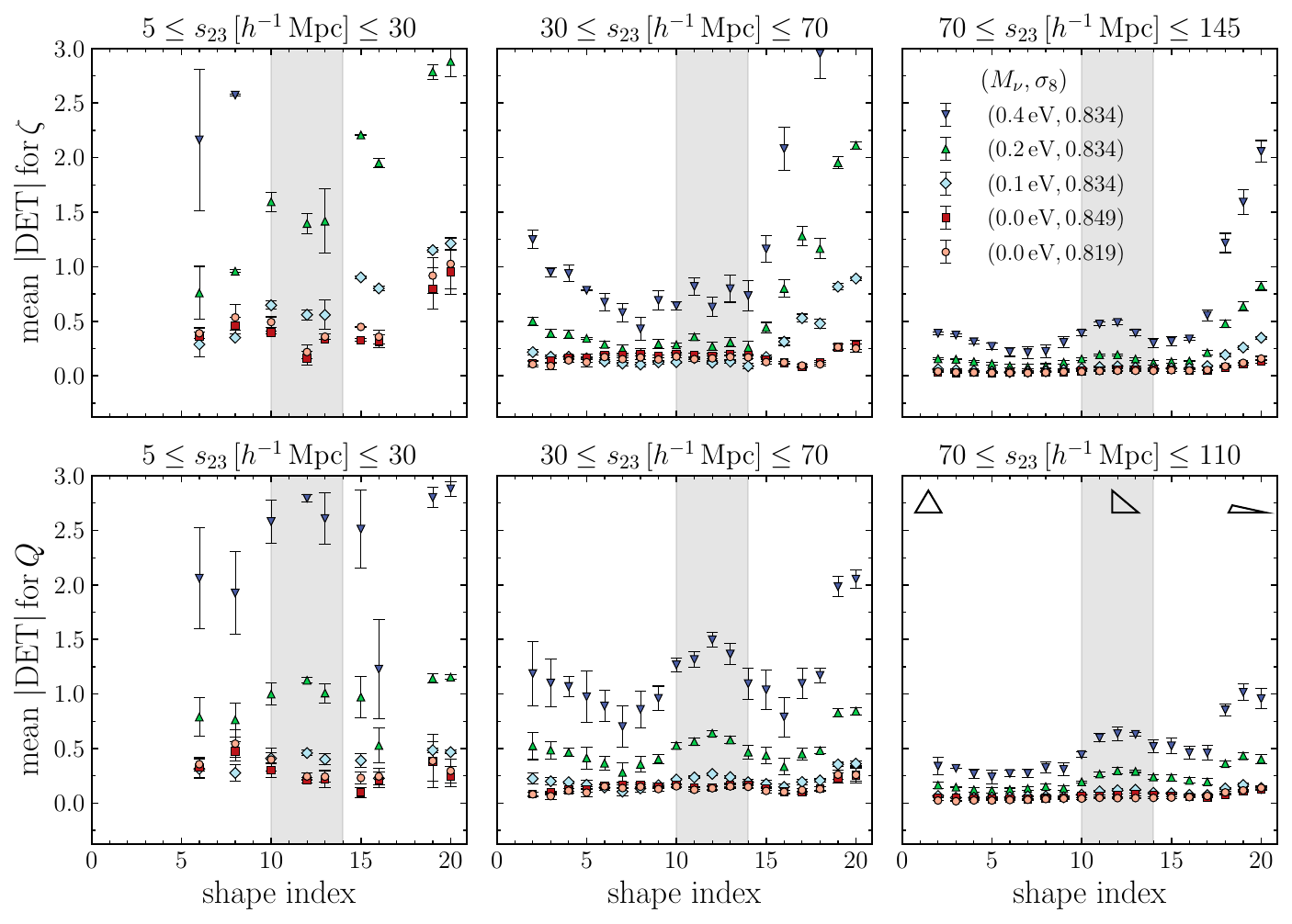}
     \caption{Mean of the absolute detectabilities (markers) as a function of the shape index (defined in Fig. \ref{fig:legend_slicing}) at $z = 0$, for the simulations with massive neutrinos and variable $\sigma_8$, and its corresponding standard error (error bars).
     The results for the connected and reduced 3PCF $\zeta$ and $Q$ are shown in the top and bottom rows, respectively. 
     Each column corresponds to a different range of triangle scales, depending on the value of the largest triangle side $s_{23}$, as detailed in the label above each panel: small ($5 \, \mpch \leq s_{23} \leq 30 \, \mpch$), intermediate ($30 \, \mpch \leq s_{23} \leq 70 \, \mpch$), and large ($70 \, \mpch \leq s_{23} \leq 145 \, \mpch$ for $\zeta$ and $70 \, \mpch \leq s_{23} \leq 110 \, \mpch$ for $Q$).
     The means are computed only for those shape indices that correspond to at least two populated $(s_{12}/s_{23}, s_{13}/s_{23})$ bins, depending on the considered scale range and the adopted binning of the side ratios. 
     The schematic representation of triangles in the bottom-right panel qualitatively shows the mapping between the shape index and triangle shapes: low values correspond to equilateral and quasi-equilateral triangles, intermediate values  (highlighted by the vertical gray band) to right-angled and quasi-right-angled triangles, and large values to elongated (squeezed/folded) triangles.
     The vertical axis shows only the mean detectability values below 3, to better highlight the differences between the lowest neutrino mass, $M_\nu = 0.1 \, \ev$, and the varying-$\sigma_8$ simulations.
     }
     \label{fig:diff_shape_det}
\end{figure*}

For each bin of shape index, we compute the mean and the standard error of the mean of the absolute values of the detectability of the connected and reduced 3PCF, for simulations with massive neutrinos as well as for those with varying $\sigma_8$. In Fig. \ref{fig:diff_shape_det}, we present the results at $z = 0$ as a function of the shape index, separating triangle configurations into small ($5 \, \mpch \leq s_{23} \leq 30 \, \mpch$), intermediate ($30 \, \mpch \leq s_{23} \leq 70 \, \mpch$), and large ($70 \, \mpch \leq s_{23} \leq s_{23,\max}$, with $s_{23,\max} = 145 \, \mpch$ for $\zeta$ and $110 \, \mpch$ for $Q$) scales. In the plot, we also show the position of particular sets of triangles, namely equilateral, right-angled (identified by the shaded region), and squeezed/folded.
In performing this analysis, we consider only those values of shape index associated with at least two populated shape bins.

On small scales, the trends appear noisier due to the smaller number of triangles averaged, compared to intermediate and large scales (respectively, 43 versus 314 and 2213 for $\zeta$, and 43 versus 314 and 842 for $Q$). The mean values for the two simulations with varying $\sigma_8$ are consistent with each other in nearly all values of the shape index, lying on average at $\sim 0.8\sigma$ from each other on intermediate and large scales, and at $\sim 1.2\sigma$ on small scales for $\zeta$ (notice, however, that the smaller number of available triangles may significantly influence this last value), and at $\sim 1\sigma$ for $Q$.
In $\zeta$, elongated shapes (high shape indices) dominate the massive neutrino signal at all scales, with a smaller yet noticeable contribution (a factor of $\sim 4$ lower) from intermediate/right-angled configurations at large scales, visible as a small bump at shape index $\sim 12$ for $M_\nu \geq 0.2 \, \ev$.
In $Q$, the bump in the neutrino signal associated with right-angled triangles is clearly visible at intermediate and large scales, and it is only slightly lower than for elongated shapes (a factor of $\sim 1.5$ smaller).

The case $M_\nu = 0.1 \, \ev$ is particularly relevant for this quantitative analysis, since for this neutrino mass the detectability values are the lowest and comparable to those of the varying-$\sigma_8$ simulations.
On small scales, the signals are difficult to distinguish within the error bars.
At intermediate and large scales, for $\zeta$ the $M_\nu$ and $\sigma_8$ signals start to be significantly distinguishable for more elongated triangles (shape index $\sim 16$); while the mean detectability values remain compatible within $\sim 2\sigma$ at lower shape indices, their differences grow much more rapidly in the massive-neutrino case, reaching discrepancies of $\sim 10\sigma$ and $\sim 7\sigma$ for elongated triangles in the ranges $30 \, \mpch \leq s_{23} \leq 70 \, \mpch$ and $70 \, \mpch \leq s_{23} \leq 145 \, \mpch$, respectively.
For $Q$, at intermediate scales the discrepancy is maximal ($\sim 5\sigma$) for shape indices $\sim 11-12$, corresponding to right-angled triangle configurations, and gets to $\sim 3\sigma$ for some elongated (shape indices $\sim 17-19$) or nearly equilateral configurations (shape indices $\sim 2-4$), while remaining $\lesssim 1\sigma$ for the remaining shapes. 
At large scales, these discrepancies reach the $\sim 3\sigma$ level for right-angled and elongated triangles and $\lesssim 2\sigma$ for nearly equilateral ones.
We warn the reader that the results for nearly equilateral configurations are based on a limited number of populated bins, as can be easily seen from Fig. \ref{fig:legend_slicing}, and may therefore be less reliable.

This analysis proves quantitatively, confirming the qualitative indication of Fig. \ref{fig:triangle_mnu_s8_deg}, that the imprint of massive neutrinos in $\zeta$ and $Q$ across different triangle shapes cannot be reproduced by a simple variation of $\sigma_8$, since at the same time, some triangle shapes react to this variation similarly to a variation in $M_\nu$, while other shapes follow a different trend.
This analysis confirms in configuration space the results obtained for the bispectrum in \citet{Hahn20}, paving the way for more systematic studies of the shape-dependent sensitivity of configuration-space higher-order statistics as a promising strategy to disentangle $M_\nu$ and $\sigma_8$-related effects.
\end{appendix}

\end{document}